\documentclass[]{emulateapj}

\usepackage{graphicx,amssymb,verbatim,times}

\newcommand\hi{\hbox{H{\scriptsize$\rm I$}}}
\newcommand\hii{\hbox{H{\scriptsize$\rm II$}}}
\newcommand\nii{\hbox{[N{\scriptsize$\rm II$}]}}
\newcommand\oiii{\hbox{[O{\scriptsize$\rm III$}]}}
\newcommand\sii{\hbox{[S{\scriptsize$\rm II$}]}}
\newcommand\oi{\hbox{[O{\scriptsize$\rm I$}]}}
\newcommand\oii{\hbox{[O{\scriptsize$\rm II$}]}}

\newcommand\lya{Ly$\alpha$}
\newcommand\ha{H$\alpha$}
\newcommand\hb{H$\beta$}

\newcommand\nicmosh{$H_{160}$}

\newcommand\civ{\hbox{C~\scriptsize$\rm IV$}}
\newcommand\ciii{\hbox{C~\scriptsize$\rm III$}}
\newcommand\heii{\hbox{He~\scriptsize$\rm II$}}
\newcommand\nv{\hbox{N~\scriptsize$\rm V $}}

\slugcomment{}

\shorttitle{The LAB B1 at $z=2.38$: another hidden quasar}
\shortauthors{R.A. Overzier et al.}

\begin{document}


\title{Resolving the optical emission lines of Ly$\alpha$\
  blob `B1' at $\mathrm{z}=2.38$: another hidden quasar}


\author{R. A. Overzier\altaffilmark{1,2},
  N. P. H. Nesvadba\altaffilmark{3}, M. Dijkstra\altaffilmark{4},
  N. A. Hatch\altaffilmark{5}, M. D. Lehnert\altaffilmark{6},
  M. Villar-Mart\'in\altaffilmark{7}, R. J. Wilman\altaffilmark{8},
  A. W. Zirm\altaffilmark{9}}
\email{overzier@astro.as.utexas.edu}

\altaffiltext{1}{Department of Astronomy, University of Texas at Austin, 1 University Station C1400, Austin, TX 78712, USA}
\altaffiltext{2}{Observat\'orio Nacional, Rua Jos\'e Cristino, 77. CEP 20921-400, S\~ao Crist\'ov\~ao, Rio de Janeiro-RJ, Brazil}
\altaffiltext{3}{Institut d'Astrophysique Spatiale, CNRS, Universit\'e Paris-Sud, 91405, Orsay, France}
\altaffiltext{4}{Max-Planck-Institut fuer Astrophysik, Karl-Schwarzschild-Str. 1, 85741, Garching, Germany}
\altaffiltext{5}{School of Physics and Astronomy, University of Nottingham, University Park, Nottingham NG7 2RD, UK}
\altaffiltext{6}{GEPI, Observatoire de Paris, UMR 8111, CNRS, Universit\'e Paris Diderot, 5 place Jules Janssen, 92190, Meudon, France}
\altaffiltext{7}{Centro de Astrobiolo\'ia (INTA-CSIC), Carretera de Ajalvir, km 4, 28850 Torrej\'on de Ardoz, Madrid, Spain}
\altaffiltext{8}{Department of Physics, University of Durham, South Road, Durham DH13LE, UK}
\altaffiltext{9}{Dark Cosmology Centre, Niels Bohr Institute, University of Copenhagen, Juliane Maries Vej 30, DK-2100 Copenhagen, Denmark}



\begin{abstract} We have used the SINFONI near-infrared integral field
  unit on the VLT to resolve the optical emission line structure of
  one of the brightest ($L_{Ly\alpha}\approx10^{44}$ erg s$^{-1}$) and
  nearest ($z\approx2.38$) of all \lya\ blobs (LABs). The target,
  known in the literature as object `B1' (Francis et al. 1996), lies
  at a redshift where the main optical emission lines are accessible
  in the observed near-infrared. We detect luminous
  \oiii$\lambda\lambda4959,5007$\AA\ and \ha\ emission with a spatial
  extent of at least $32\times40$ kpc ($4\arcsec\times5\arcsec$). The
  dominant optical emission line component shows relatively broad
  lines (600--800 km s$^{-1}$, FWHM) and line ratios consistent with
  AGN-photoionization.  The new evidence for AGN photo-ionization,
  combined with previously detected \civ\ and luminous, warm infrared
  emission, suggest that B1 is the site of a hidden quasar. This is
  confirmed by the fact that \oii\ is relatively weak compared to
  \oiii\ (extinction-corrected \oiii/\oii\ of about 3.8), which is
  indicative of a high, Seyfert-like ionization parameter. From the
  extinction-corrected \oiii\ luminosity we infer a bolometric AGN
  luminosity of $\sim3\times10^{46}$ erg s$^{-1}$, and further
  conclude that the obscured AGN may be Compton-thick given existing
  X-ray limits.  The large line widths observed are consistent with
  clouds moving within the narrow line region of a luminous QSO. The
  AGN scenario is capable of producing sufficient ionizing photons to
  power the \lya, even in the presence of dust. By performing a census
  of similar objects in the literature, we find that virtually all
  luminous LABs harbor obscured quasars. Based on simple duty-cycle
  arguments, we conclude that AGN are the main drivers of the \lya\ in
  LABs rather than the gravitational heating and subsequent cooling
  suggested by cold stream models. We also conclude that the empirical
  relation between LABs and overdense environments at high redshift
  must be due to a more fundamental correlation between AGN (or
  massive galaxies) and environment.
\end{abstract}




\section{Introduction}
\label{sec:intro}

The nature of the large, spatially extended regions of luminous line
emission found around many types of active galactic nuclei (AGN) at
both low and high redshifts (e.g., radio galaxies, quasars and
Seyferts) have been studied for over three decades
\citep[e.g.][]{heckman82,tadhunter86,baum89,mccarthy89,fu09}. More
recently, qualitatively similar structures were found also toward
other lines of sight that are not (or, at least, not obviously)
associated with luminous AGN
\citep[e.g.][]{francis96,steidel00,prescott08,erb11}. These nebulae,
often referred to as \lya\ halos or ``blobs'', are most conspicuous at
the redshifted wavelength of \lya, reaching sizes of order 100 kpc and
line luminosities of $\sim10^{44}$ erg s$^{-1}$ (up to $\sim200$ kpc
and $\sim10^{45}$ erg s$^{-1}$ for some high redshift radio galaxies,
HzRGs).

For the origin of the \lya\ emitting gas as well as its main source of
ionization there are as many theories as there are \lya\ blobs
(LABs). Their rarity and association with large amounts of warm,
ionized gas suggests a link with massive galaxy assembly, with the
\lya\ either tracing the giant gas reservoir from which the galaxy is
being formed, or the gas that has been expelled by the subsequent
superwind from a starburst or AGN. Many radio-quiet LABs show evidence
for the presence of an AGN
\citep[e.g.][]{francis96,basu-zych04,chapman04,dey05,scarlata09,geach09,yang09,colbert11},
as well as obscured starbursts \citep{geach05,colbert11} and
outflowing superwinds \citep[e.g.][]{bower04,wilman05,weijmans10}. The
LAB phenomenon has also been connected to the popular `cold flow'
model of galaxy formation, in which the \lya\ is largely powered by
collisionally excited \hi\ in filamentary streams leading to the
object \citep[e.g.  ][]{fardal01,dijkstra09}. While these predictions
are consistent with some of the main blob phenomenology such as their
luminosity (function) and morphologies, these models are difficult to
test empirically in the presence of heavy star-formation, dust,
outflows, merging, or AGN. Also, some LABs show relatively high
metallicities or metallicity gradients, indicating that at least the
central gas has already been enriched \citep{francis96,overzier01}.
\begin{figure*}[t]
\begin{center}
\includegraphics[height=5cm]{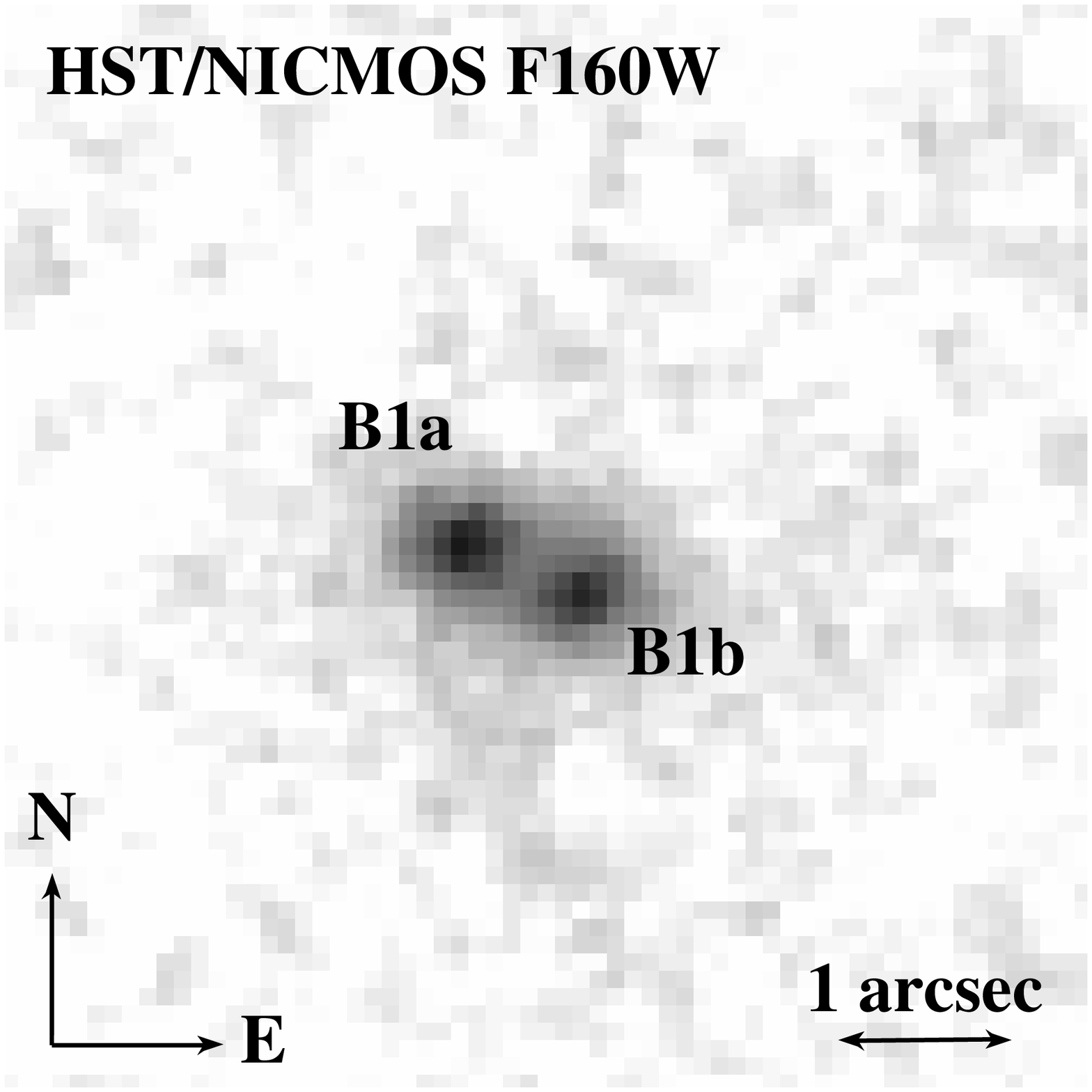}
\includegraphics[height=5cm]{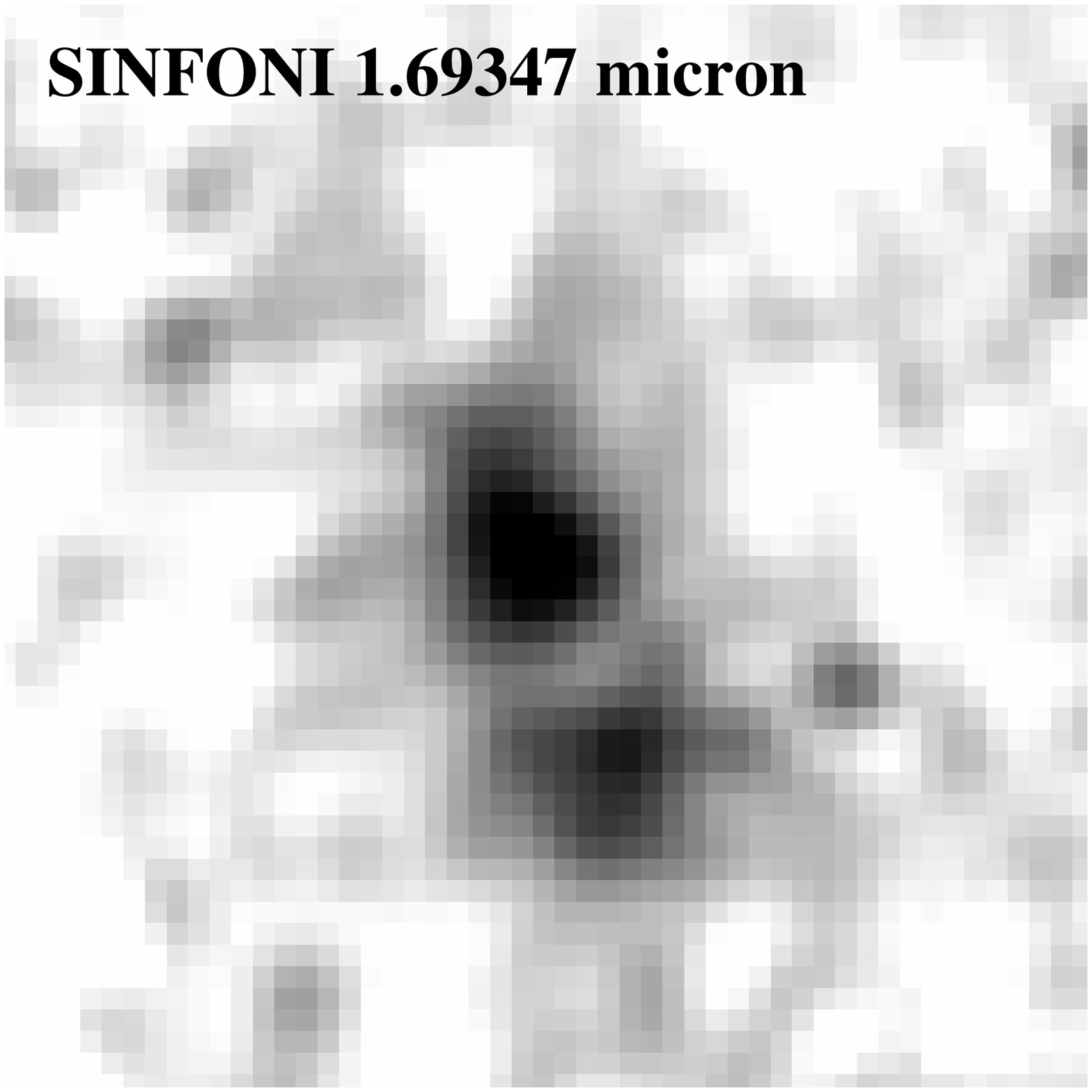}
\includegraphics[height=5cm]{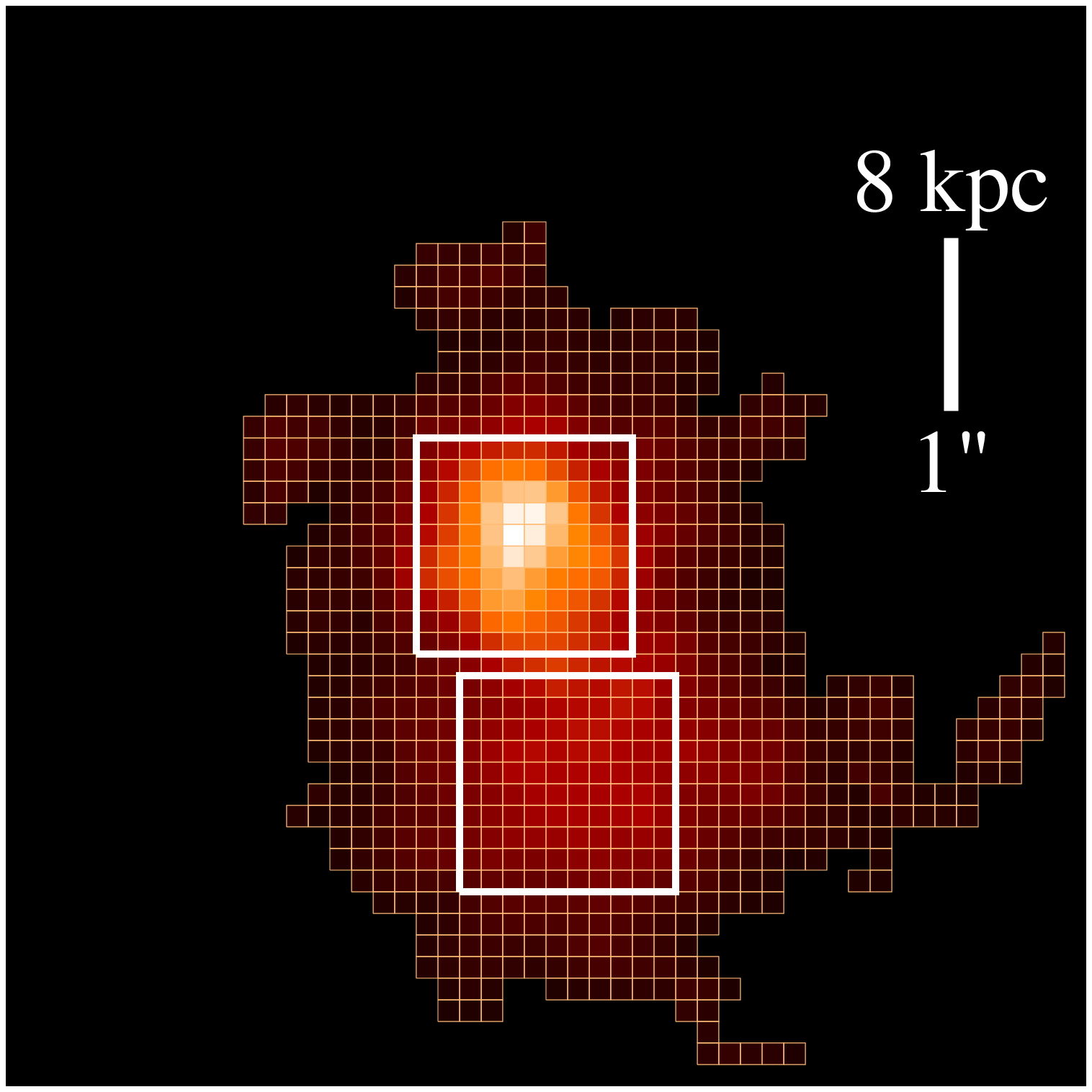}
\end{center}
\caption{\label{fig:images}$H$-band continuum and
  \oiii$\lambda$5007\AA\ emission line morphologies associated with
  the \lya\ blob B1 at $z=2.38$. Panels from left to right show the
  registered HST/NICMOS \nicmosh\ image, the SINFONI $H$-band channel
  map centered at the peak of the \oiii$\lambda$5007\AA\ emission
  line, and a map of the \oiii\ emission line flux detected at
  $>3\sigma$. The maps are $6\farcs25\times6\farcs25$ or $50\times50$
  kpc across. The two square extraction apertures centered on the
  brightest emission line regions B1 North and South discussed in this
  paper are indicated in the right-hand panel. The two red galaxies
  `B1a' and `B1b' identified by \citet{francis01} roughly coincide
  with the location of B1 North, but we note that the registration of
  the HST and SINFONI maps is uncertain by $\sim$0\farcs5 due to the
  lack of strong continuum in the SINFONI data.}
\end{figure*}

In the \lya\ halos associated with radio galaxies, the \lya\ line
emission is predominantly powered by photo-ionization from an AGN,
often with a smaller contribution from shock waves associated with
radio jets \citep[e.g.][]{humphrey08}. The main arguments in favor of
AGN photoionization are the biconical morphologies of the emission
line gas observed in some sources, the energetics and line ratios of
the extended emission, and the presence of emission line gas well
beyond the radio structure
\citep[e.g.][]{mccarthy90,villar-martin97,villar-martin02,villar-martin03,best00,humphrey08}. In
addition to this, shocks are often invoked to explain the highly
perturbed gas motions, shock-like line ratios, high gas temperatures,
and energetics of the extended emission most closely associated with
the radio structures or in regions well beyond the photoionizing
volume
\citep[e.g.][]{dopita95,villar-martin99,solorzano01,nesvadba07a,humphrey08}.
A fraction of HzRGs at $z\gtrsim2$ have exceptionally luminous \lya\
halos best explained by an additional source of ionizing photons
(besides the AGN) that must originate from star formation taking place
on scales of tens of kpc throughout the halo
\citep[e.g.][]{pentericci98,villar-martin07a,hatch08}. These
starbursts sometimes produce powerful outflows that eject energy into
the surrounding gas \citep{zirm05}, analogous to the strong
starburst-driven outflows observed in some LABs.

In this Paper we study the LAB J2143--4423 (`B1') at $z=2.38$
discovered by \citet{francis96}. B1 has a \lya\ luminosity of
$\sim8\times10^{43}$ erg s$^{-1}$ \citep{francis96,palunas04}. B1 is
also a luminous mid-IR source ($\sim0.24$ mJy at 24 $\mu$m), possibly
with polycyclic aromatic hydrocarbon (PAH) emission detected at 7.7
$\mu$m and substantial underlying continuum
\citep{colbert06,colbert11}. This may suggest that B1 harbors an
obscured starburst (star formation rate of $\sim$420 $M_\odot$
yr$^{-1}$ ), as well as an obscured AGN \citep{colbert11}. The object
is radio-quiet with a (3$\sigma$) upper limit on its radio flux of 3.3
mJy \citep{francis96}. \citet{francis01} identified two red galaxies
separated by a projected $\sim$8 kpc near the peak\footnote{The exact
  location of the peak \lya\ emission with respect to the galaxies is
  hard to determine as the \lya\ centroid changes as a function of
  resolution and smoothing scale. If measured at arcsec resolution the
  peak lies about 1\arcsec\ South of the galaxies, while it moves
  North at larger smoothing scales. At HST resolution, a \lya\ point
  source is detected coinciding with the easternmost galaxy, but this
  point source only contains $\sim13$\% of the total \lya\ flux of B1
  \citep{francis96,francis01}.} of the \lya\ emission of B1 (see the
left panel of Fig. \ref{fig:images}). Lower surface brightness \lya\
emission appears to follow a faint rest-frame UV filament that extends
a few arcseconds to the South of the galaxy pair. Although the source
is not detected in the X-rays \citep[see][]{colbert11}, it has a
significant detection\footnote{The exact location of the \civ\
  emission is uncertain. \citet{francis96} state that \civ\ coincides
  with the peak of \lya, while \citet{francis12} argue that the \civ\
  comes from the `southern end of B1' on the basis of the same 1996
  data.} in \civ\
\citep[$F_{Ly\alpha}/F_{CIV}\approx7$,][]{francis96}.  An
interpretation of the entire system was given by \citet{francis12},
who suggested that the \lya\ (and \civ) emission in B1 could be
powered by shocks produced at the interface between a hot central
medium and numerous infalling cold clouds, disfavoring the AGN
interpretation.

Here, we present an analysis of observations of the main rest-frame
optical emission of B1, performed with the Spectrograph for Integral
Field Observations in the Near Infrared (SINFONI). The structure of
the Paper is as follows. We first describe our data and methods of
analysis (Section 2). We then perform an analysis of the
two-dimensional and integrated spectra (Section 3), followed by a
discussion of the results (Section 4). We use a cosmology in which the
angular scale at $z=2.38$ amounts to 8.0 kpc arcsec$^{-1}$ ($H_0=73$
km s$^{-1}$ Mpc$^{-1}$, $\Omega_m=0.27$, $\Omega_\Lambda=0.73$).

\begin{figure*}[t]
\begin{center}
\includegraphics[width=0.25\textwidth]{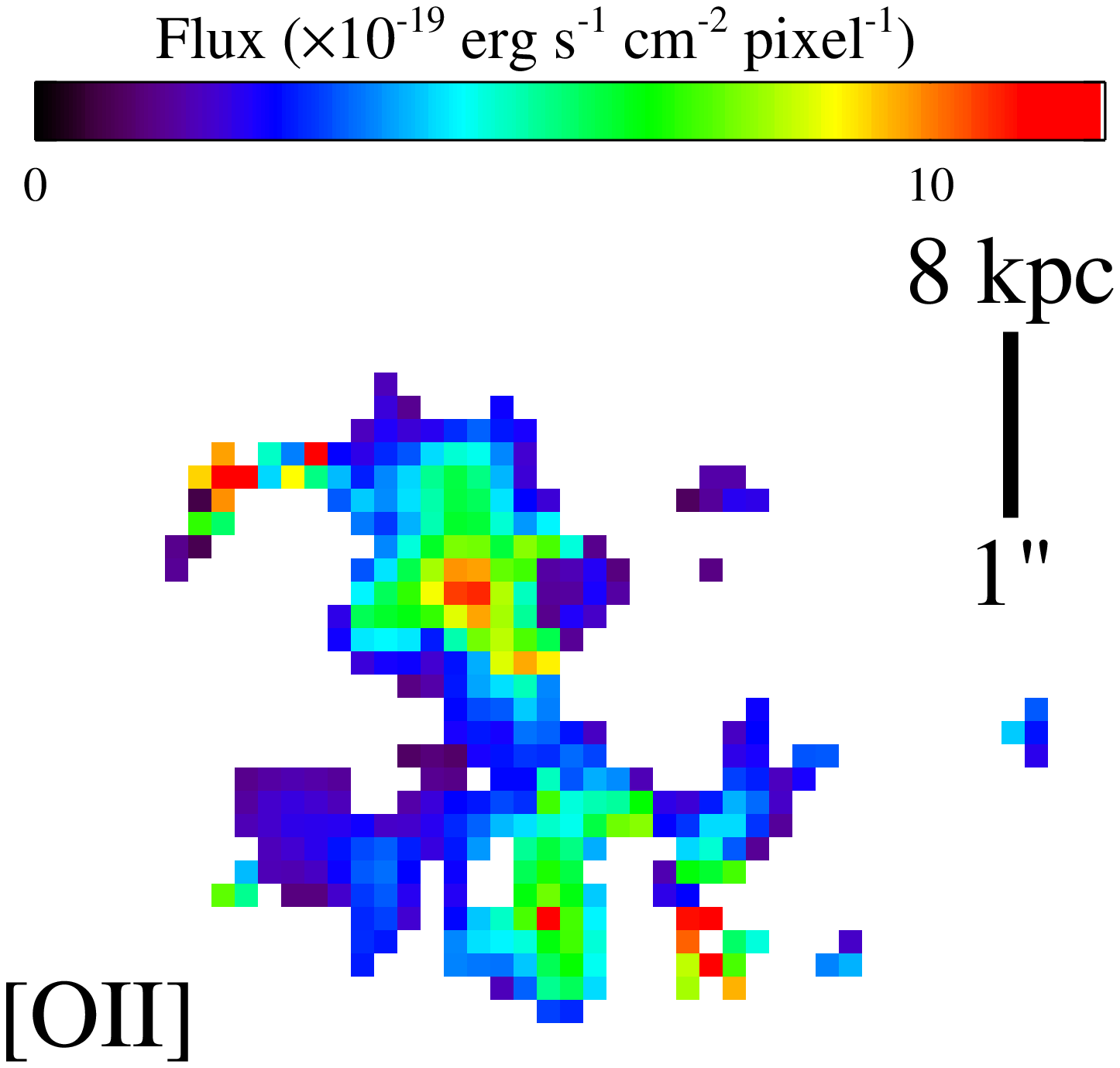}
\includegraphics[width=0.25\textwidth]{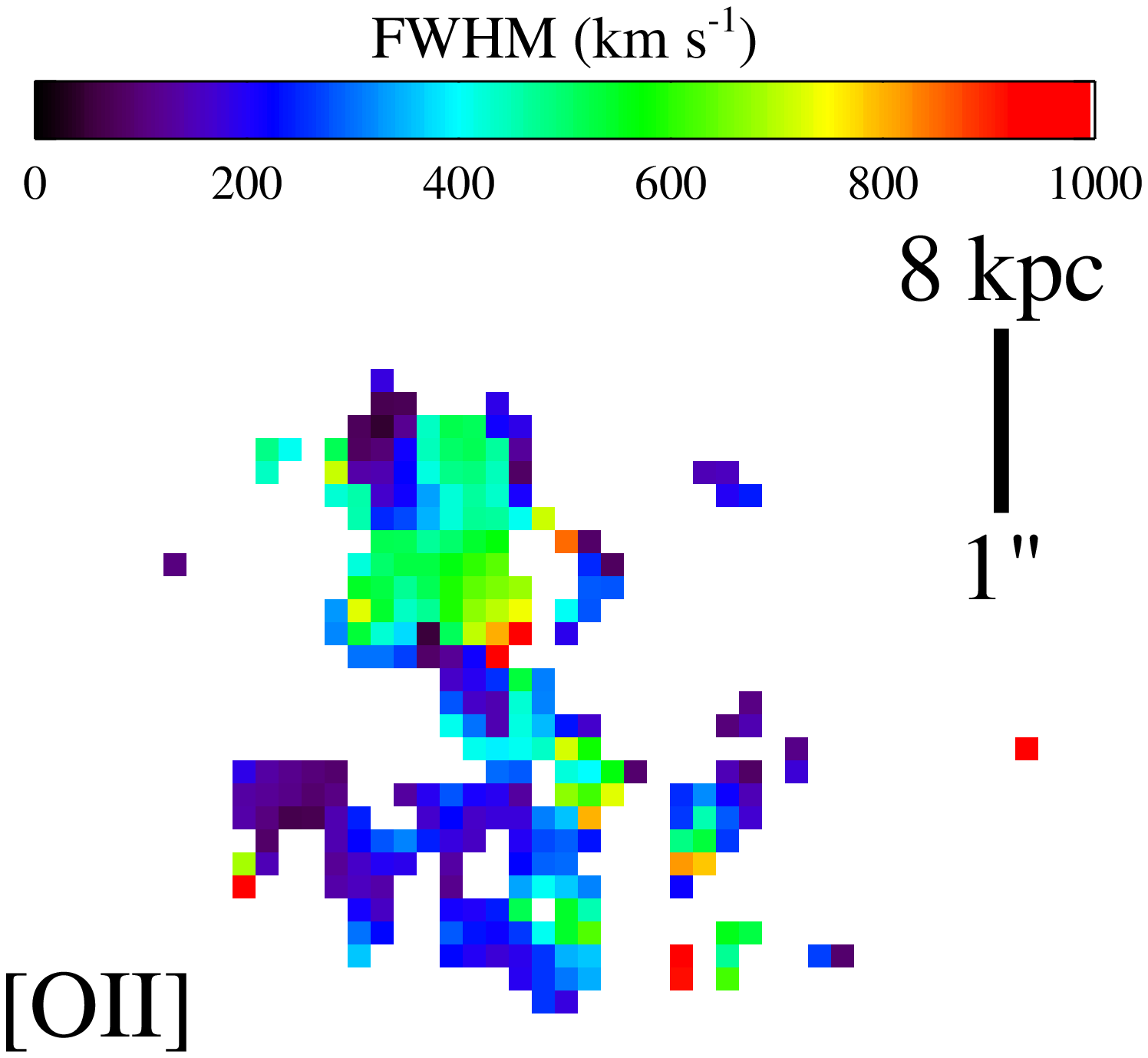}
\includegraphics[width=0.25\textwidth]{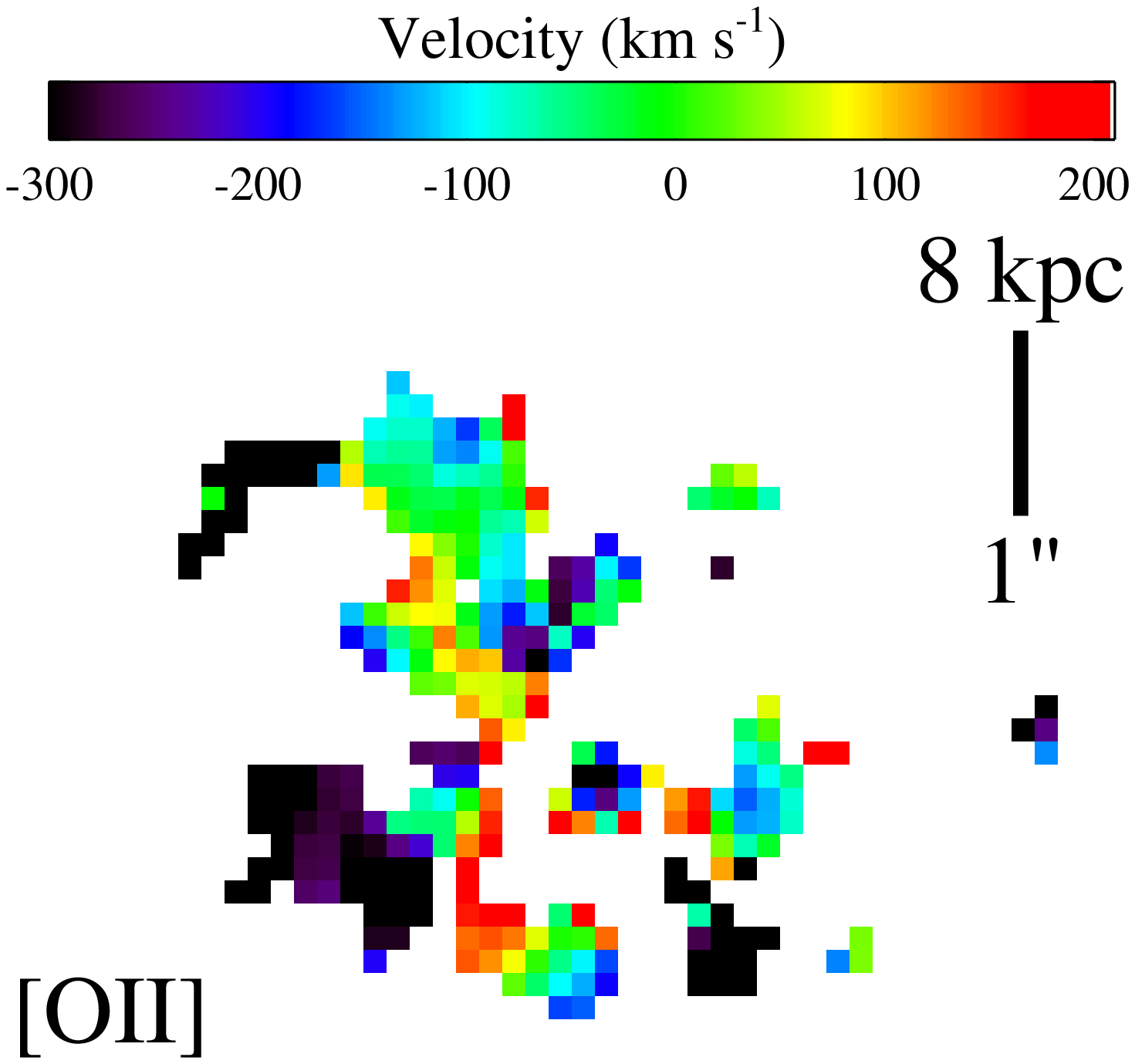}\\
\vspace{10mm}
\includegraphics[width=0.25\textwidth]{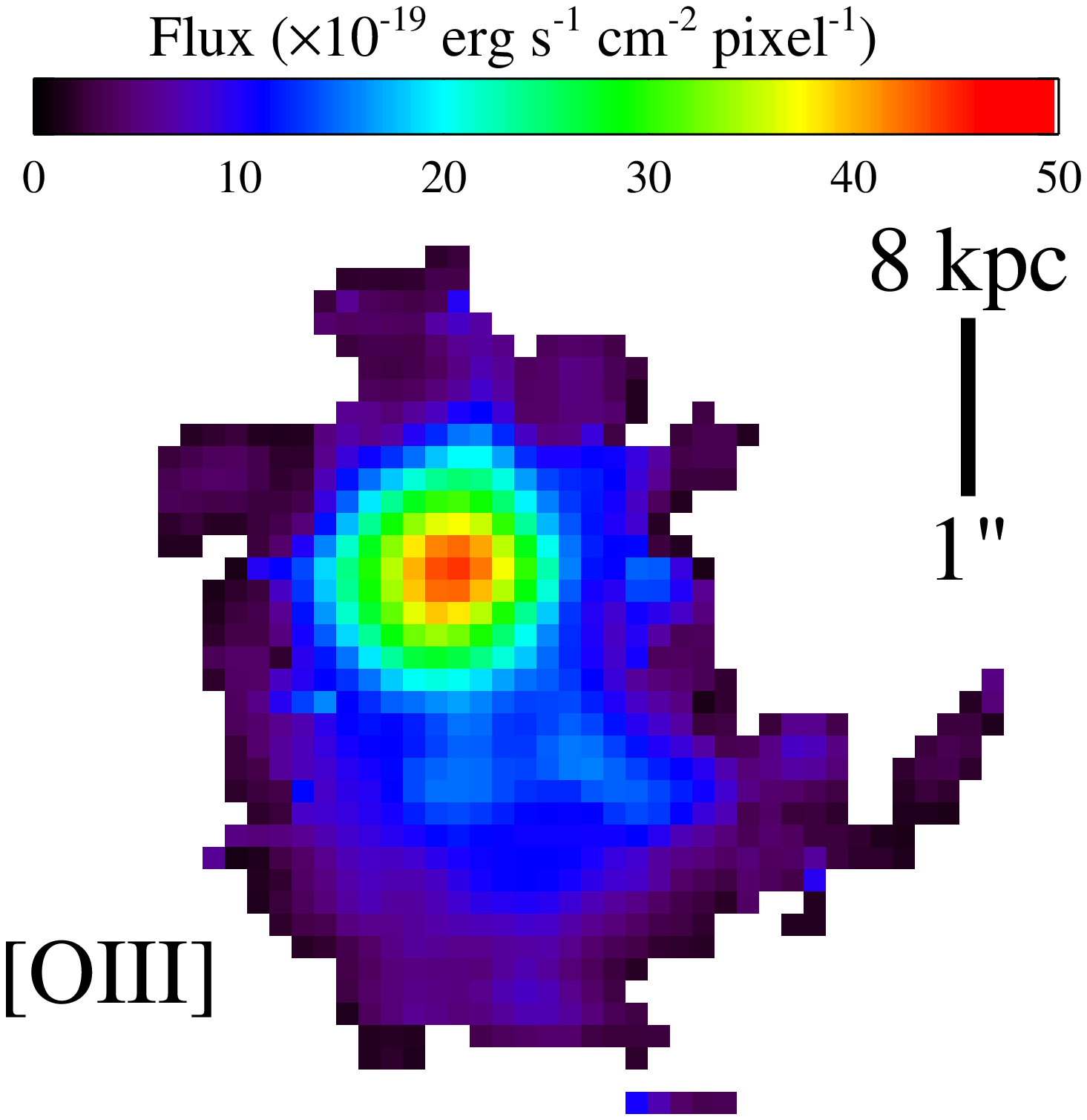}
\includegraphics[width=0.25\textwidth]{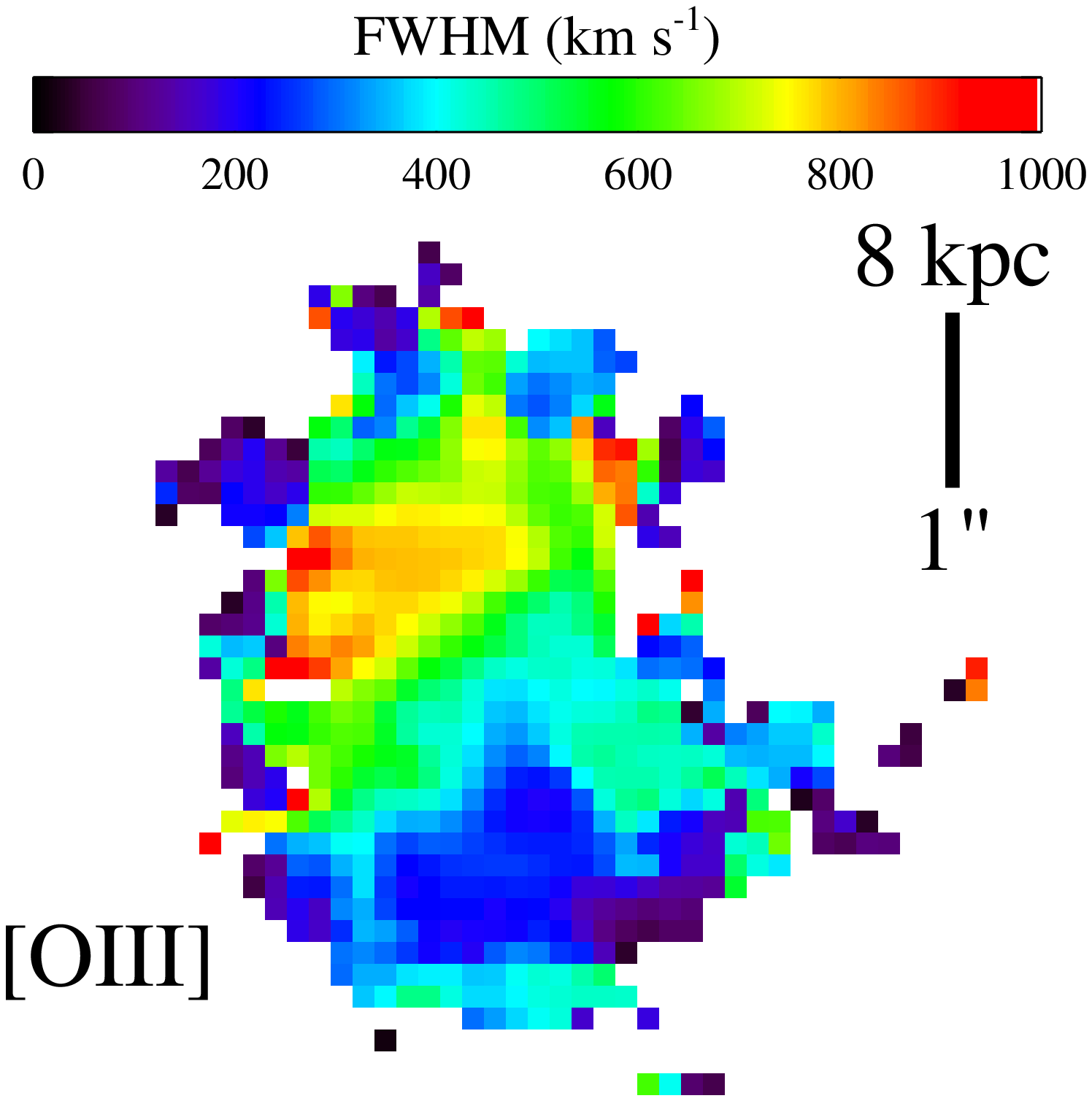}
\includegraphics[width=0.25\textwidth]{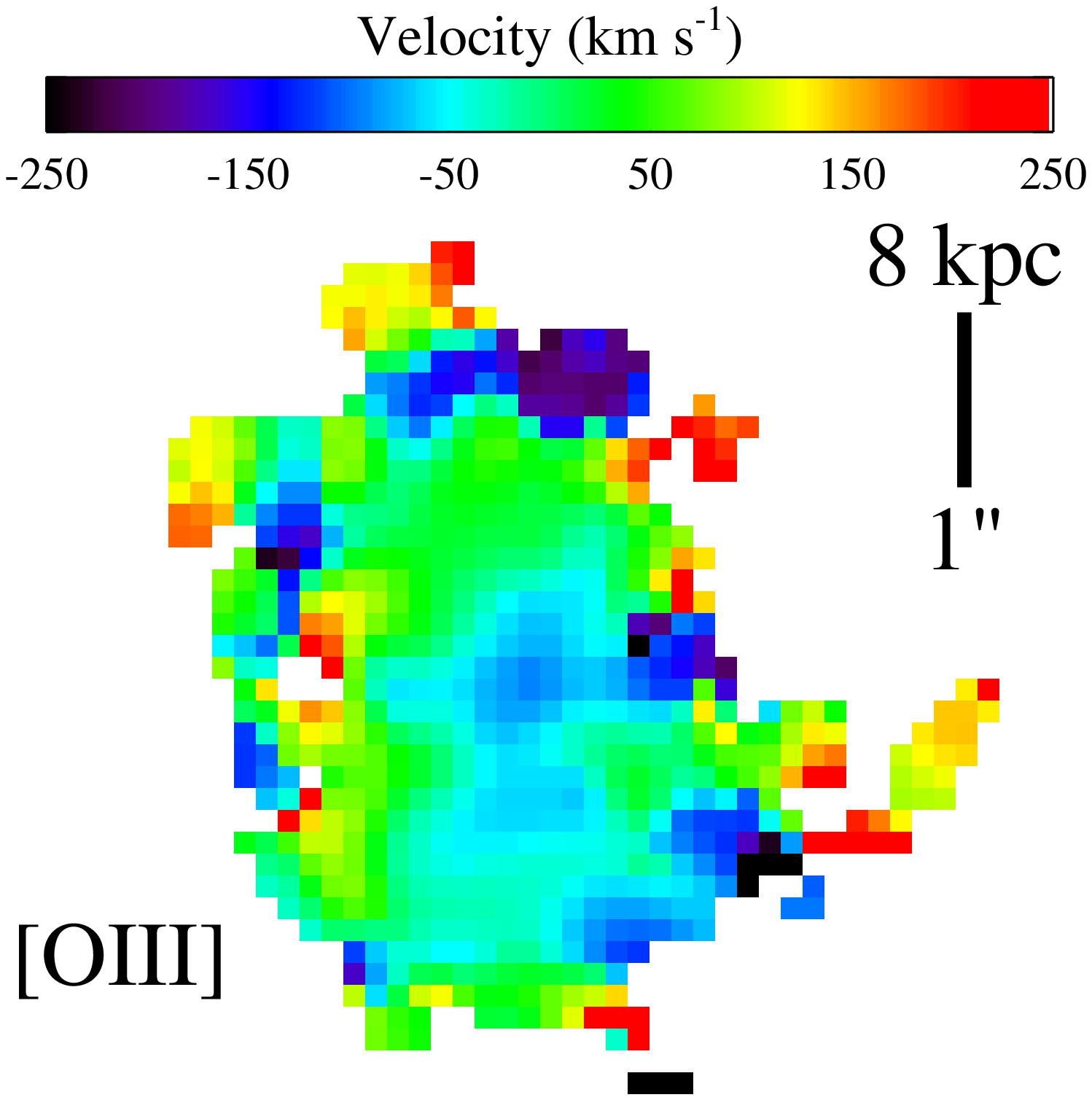}\\
\vspace{10mm}
\includegraphics[width=0.25\textwidth]{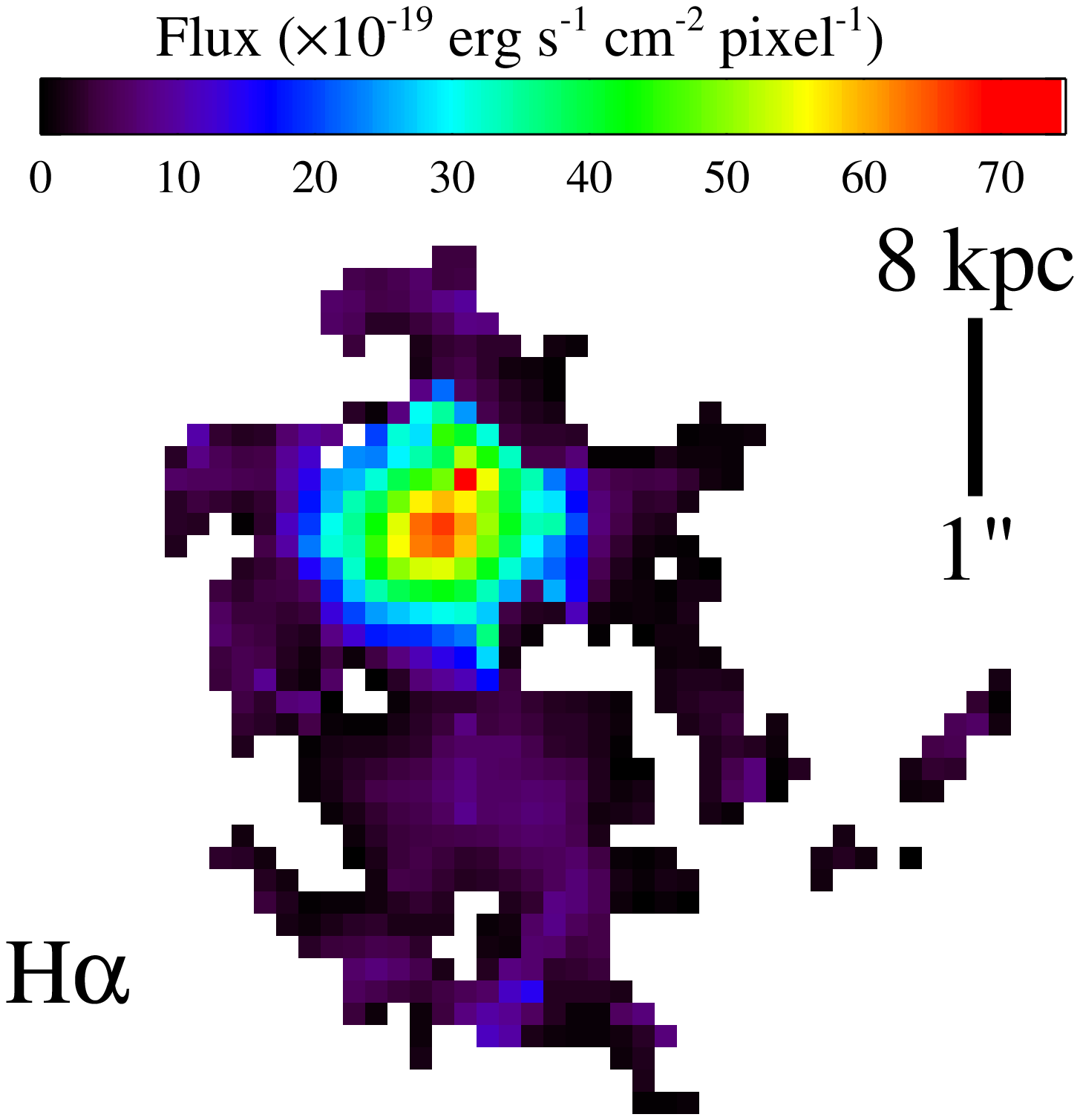}
\includegraphics[width=0.25\textwidth]{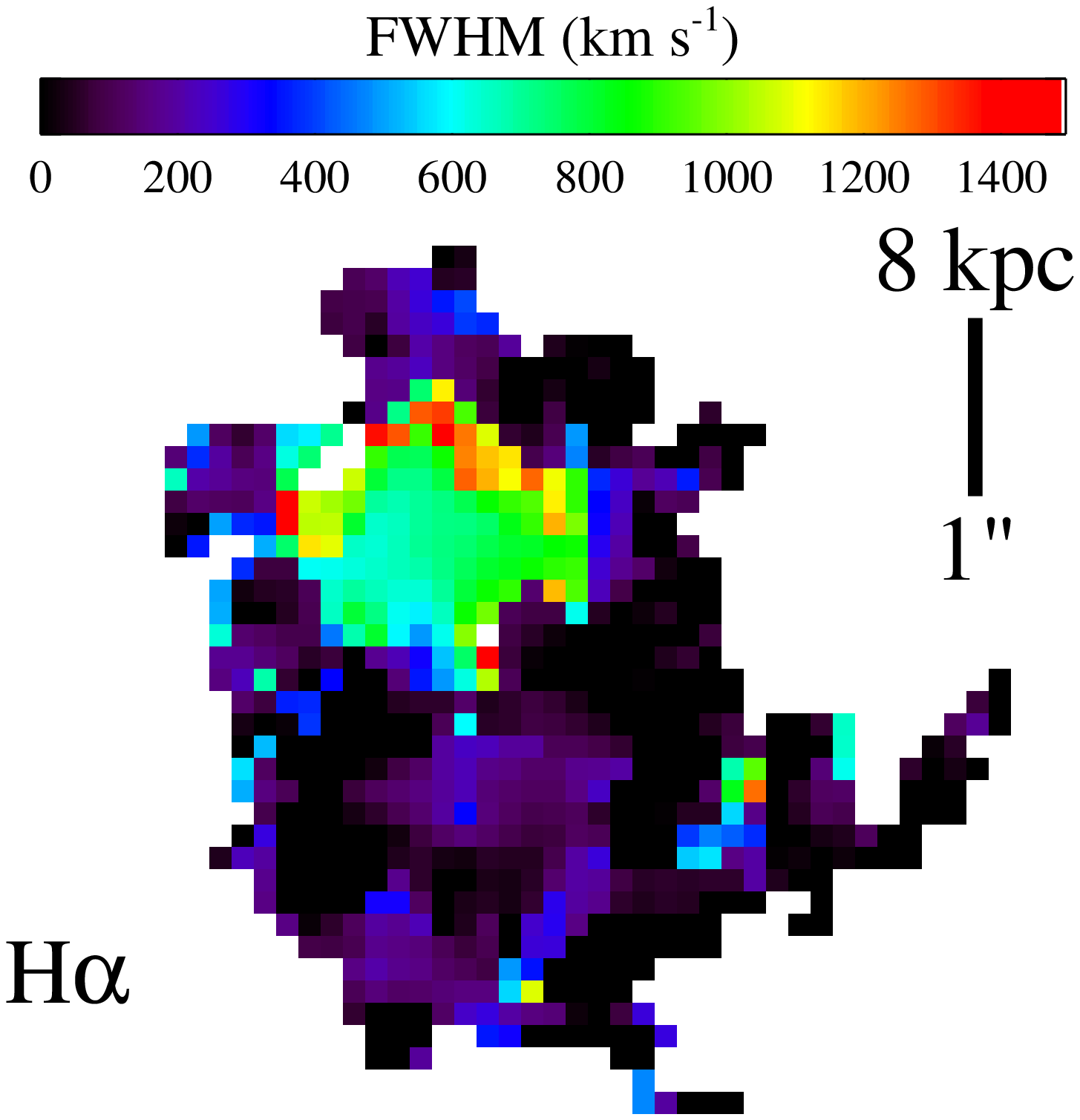}
\includegraphics[width=0.25\textwidth]{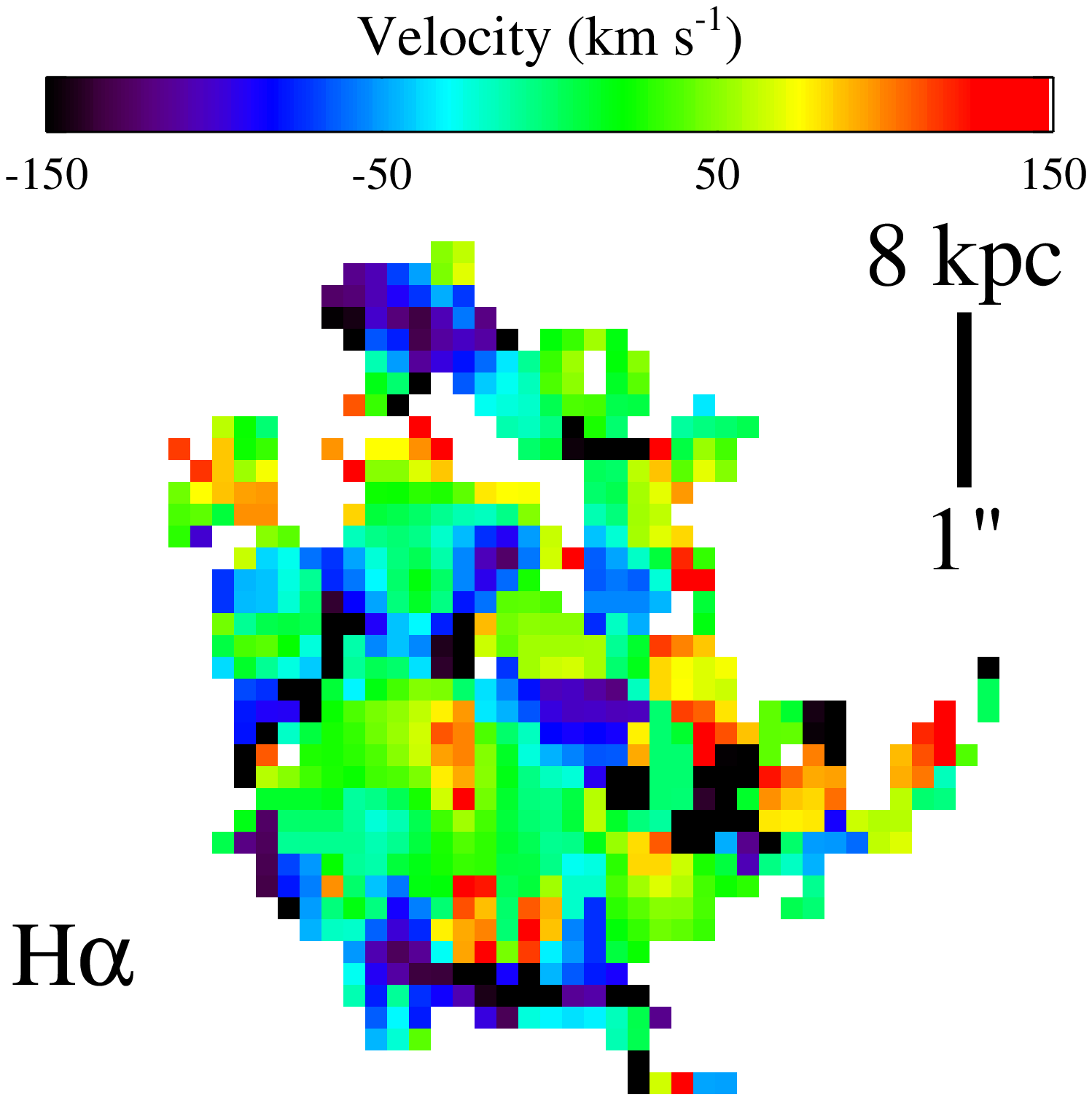}
\end{center}
\caption{\label{fig:maps} Spatially resolved maps of the dominant
  emission lines: \oii$\lambda\lambda$3726,3729\AA\ (top),
  \oiii$\lambda$5007\AA\ (middle), and \ha\ (bottom). Panels from left
  to right show the line flux {\it (a)}, the FWHM velocity width {\it
    (b)}, and the line velocity relative to that at the peak of the
  \oiii\ emission {\it (c)}.}
\end{figure*}

\begin{figure*}[t]
\begin{center}
 \includegraphics[width=0.4\textwidth]{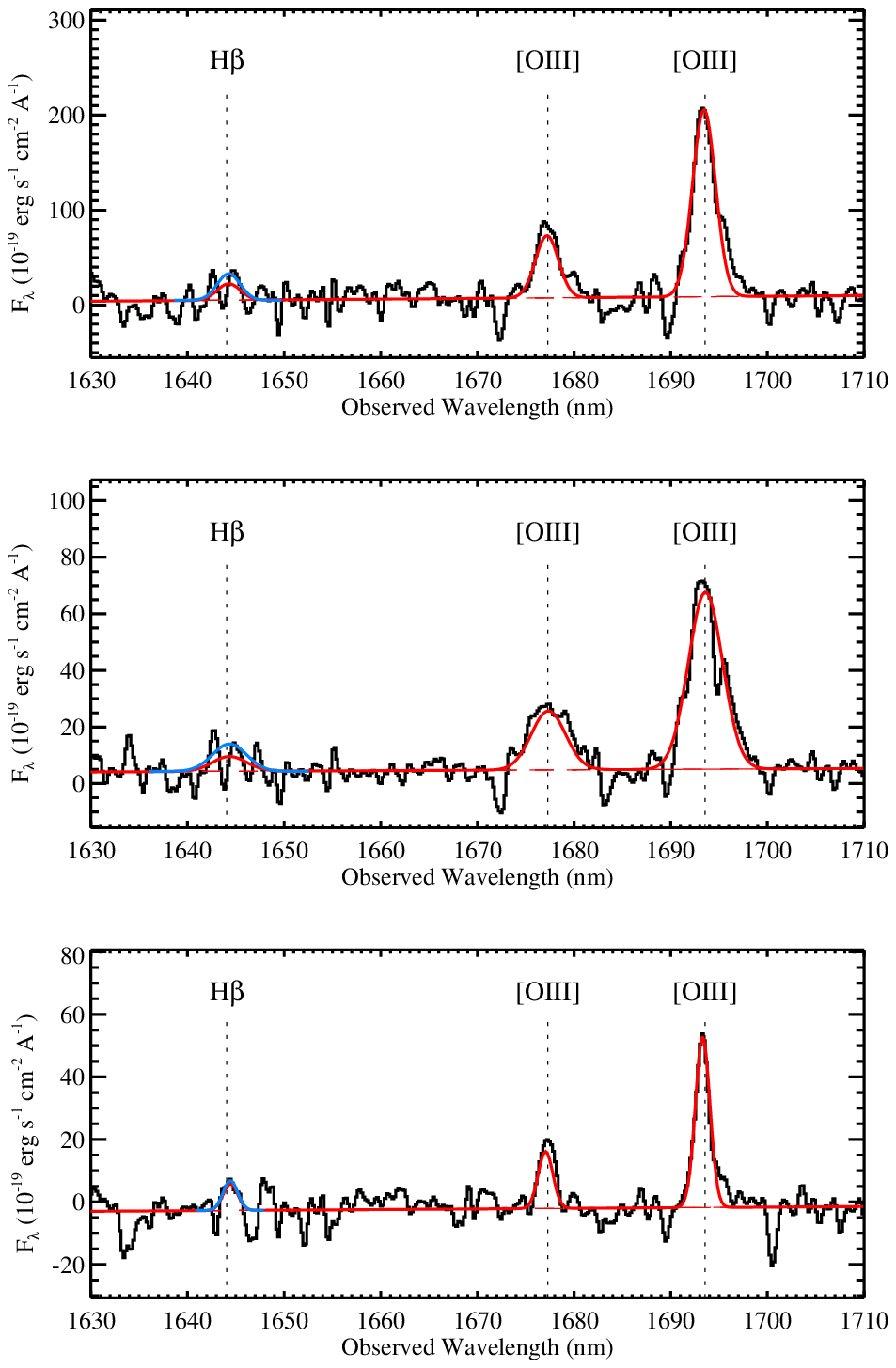}
 \includegraphics[width=0.4\textwidth]{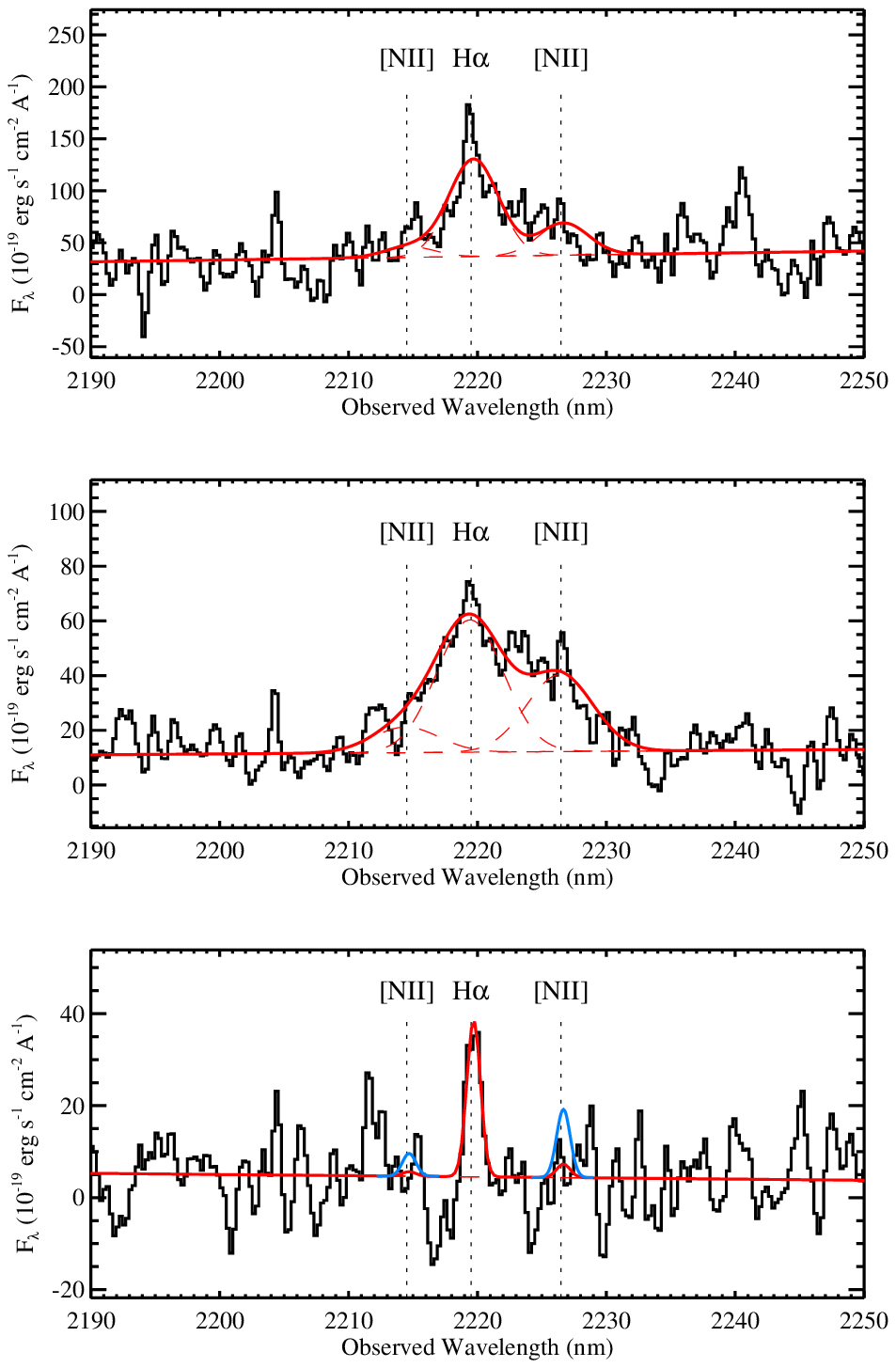}
\end{center}
\caption{\label{fig:spectra}Integrated spectra extracted from 3
  different regions of the B1 halo. Panels on the top row correspond
  to all $S/N>5$ pixels within the masked region shown in the right
  panel of Fig. \ref{fig:images}, while the middle and lower panels show
  the spectra extracted in the `North' and `South' regions of B1
  (white boxes in right panel of Fig. \ref{fig:images}),
  respectively. Panels on the left show the
  \oiii$\lambda\lambda$4959\AA,5007\AA\ line doublet in $H$, while
  panels on the right show the spectral region centered on \ha\ and
  the \nii$\lambda\lambda$6548\AA,6584\AA\ doublet in the
  $K$-band. Vertical lines indicate the observer-frame wavelengths of
  the lines at $z=2.381$. The \oiii\ and \ha\ profiles appear
  significantly broadened in B1 North (middle panels) with respect to
  B1 South (bottom panels).  The spectra were convolved with a
  Gaussian profile having a FWHM width equal to the instrumental
  resolution in the corresponding band. Gaussian line fits are
  indicated in red. $2\sigma$ upper limits on \hb\ and \nii\ are
  indicated in blue. }
\end{figure*}

\section{Data and Analysis Methods}

We have observed B1 using SINFONI on the VLT in service mode\footnote{Proposal ID:
  081.A--0604(A)} in July 2008. We used the seeing-limited mode giving a pixel scale of
$0\farcs125\times0\farcs125$ and a field of view of
8\arcsec$\times$8\arcsec. Because the lines are faint, sky or `off'
frames taken only once per 1 hr Observation Block (OB) were scaled and
subtracted using the object-free regions of the dithered science or
`on' frames. We obtained 24/6 `on'/`off' frames of 600 s each in $J$
and $K$, and 43/13 `on'/`off' frames of 300 s each in $H$. The total
science exposure time was 4 hr in each of $J$ and $K$, and 3.6 hr in
$H$.  The data are calibrated following the procedures described in
\citet{nesvadba08}. In brief, the frames are corrected for dark
current and flat-fielded. The absolute positions of each slitlet are
determined based on the standard SINFONI calibration data. The curved
spectra are rectified and wavelength calibrated based on arc
spectra. At each wavelength, the sky frames are subtracted from the
science frames after normalizing the sky to the average measured in
the object frame. For each OB, the calibrated frames are spatially
aligned based on the astrometry recorded in the headers, and a final
combined data cube is constructed after aligning the data from each OB
based on a cross-correlation of the line images. We perform a telluric
correction and a flux-calibration based on standard star observations.

In order to limit the effects from the sometimes substantial sky
residuals, some sigma clipping was performed in post-processing. An
object mask was generated by fitting the main emission lines (\oiii\
$\lambda$5007 in $H$ and \ha\ in $K$) and masking out each pixel with
a $<3\sigma$ detection. Low and high values were rejected and the mean
value of the unmasked region was used for an additional background
subtraction. In the analysis below, before fitting individual pixels
the data cubes were convolved with a $3\times3$ (FWHM) pixel Gaussian
along the spatial axes, and with a Gaussian filter having a FWHM equal
to the instrumental resolution (approximately 6.7\AA\ in $J$, 5.7\AA\
in $H$, and 4.9\AA\ in $K$) along the dispersion axis. Integrated
spectra extracted from larger regions were only convolved along the
dispersion axis. The \oii$\lambda\lambda$3726,3729 line doublet in $J$
was fitted as a single Gaussian emission line, based on
signal-to-noise (S/N) and spectral resolution considerations. Groups
of emission lines (\hb\ and \oiii$\lambda\lambda$4959,5007 in $H$, and
\ha\ and \nii$\lambda\lambda$6548,6584 in $K$) were fitted using a
series of single Gaussians plus a continuum. The flux ratios of the
\oiii\ and \nii\ line doublets were fixed to the expected 1:3 ratio,
and their widths were forced to be equal. The final object mask based
on all $>3\sigma$ detections of the brightest and cleanest line
(\oiii\ $\lambda$5007) is shown in the right panel of
Fig. \ref{fig:images}.

\begin{table}[t]
\begin{center}
\caption[]{\label{tab:lines}Emission line measurements of B1 North and South}
\begin{tabular}{lccc}
\hline
\hline
Line          & $z_{cen}$ & Flux$^{a}$  & FWHM$^{b}$ \\
\hline
\multicolumn{4}{l}{\bf B1 North}\\
\oii$\lambda$3727$^c$ & $2.3839\pm0.0004^c$ & $0.5\pm0.04^c$  & $786\pm79^c$ \\         
\hb\ &  \multicolumn{1}{c}{--} & \multicolumn{1}{c}{$<0.5^d$}  & \multicolumn{1}{c}{--}\\
\oiii$\lambda$4959 & $2.3811\pm0.0001$ & $0.9\pm0.03$  & $724\pm21$ \\         
\oiii$\lambda$5007 & $2.3811\pm0.0001$ & $2.8\pm0.1$   & $724\pm21$  \\
\ha & $2.3809\pm0.0002$ & $3.1\pm0.14$   & $816\pm44$    \\
\nii$\lambda$6549  & $2.3809\pm0.0002$ & $0.6\pm0.04$ & $816\pm44$   \\     
\nii$\lambda$6584  & $2.3809\pm0.0002$ & $1.8\pm0.12$   & $816\pm44$ \\
\sii$\lambda\lambda$6718,6732 & $2.3812\pm0.0005$ & $0.7\pm0.2$ & $400\pm130$\\
\hline
\multicolumn{4}{l}{\bf B1 South}\\
\oii$\lambda$3727$^c$ & $2.3849\pm0.0004^c$ & $0.4\pm0.04^c$  & $595\pm81^c$ \\         
\hb\ &  \multicolumn{1}{c}{--} & \multicolumn{1}{c}{$<0.2^d$} & \multicolumn{1}{c}{--}\\
\oiii$\lambda$4959 & $2.3805\pm0.0001$ &$0.4\pm0.03$ &  $281\pm15$ \\         
\oiii$\lambda$5007 & $2.3805\pm0.0001$ &$1.0\pm0.04$ & $281\pm15$   \\
\ha & $2.3813\pm0.0001$ & $0.5\pm0.07$ &$140\pm34$    \\
\nii$\lambda$6549  & \multicolumn{1}{c}{--} &
\multicolumn{1}{c}{$<0.07^d$} &  \multicolumn{1}{c}{--} \\     
\nii$\lambda$6584 & \multicolumn{1}{c}{--} & \multicolumn{1}{c}{$<0.2^e$} &  \multicolumn{1}{c}{--}\\
\hline
\hline
\end{tabular}
\end{center}
\begin{scriptsize}
  $^a$ Flux is given in units of 10$^{-16}$ erg s$^{-1}$ cm$^{-2}$.\\
  $^b$ FWHM is given in units of km s$^{-1}$.\\
  $^c$ The redshift, flux, and FWHM values given here are those obtained by fitting the \oii$\lambda\lambda$3726,3729 doublet with a single Gaussian line. The FWHM should be divided by $\sqrt{2}$ in order to get the intrinsic velocity dispersion.\\
  $^d$ A $2\sigma$ upper limit on the \hb\ flux is calculated by
  assuming that \hb\ has the same line width as \oiii\ and a peak
  height equal to $2\times\mathrm{rms}$ measured in the
  spectral region near \hb.\\
  $^e$ A $2\sigma$ upper limit on the \nii\ $\lambda$6584 flux is
  calculated by assuming that \nii\ has the same line width as \ha\
  and a peak height equal to $2\times\mathrm{rms}$ measured in the
  spectral region near \nii\ $\lambda$6584. The corresponding upper
  limit on \nii\ $\lambda$6549 is calculated by assuming that
  the line flux is 1/3 of that of \nii\ $\lambda$6584.\\
\end{scriptsize}
\end{table}

\section{Results}
\label{sec:fluxes}

\subsection{Measurements}

We obtained good, spatially resolved detections of \oii\ in $J$,
\oiii\ in $H$, and the \ha\ and \nii\ line complex in $K$. We also
obtained a possible detection in $K$ of the \sii\
$\lambda\lambda$6716,6731\AA\ doublet, but we note there are
significant sky residuals in this part of the spectrum that complicate
the analysis. In the middle panel of Fig. \ref{fig:images} we show a
1.693 $\mu$m channel map indicating the morphology of the \oiii\
$\lambda$5007\AA\ line relative to the continuum morphology from HST
(left panel). Although the absolute registration of the SINFONI data
with the images proved somewhat problematic, the region of strongest
\oiii\ emission roughly coincides with the location of the two
galaxies detected with NICMOS. Because these galaxies are
approximately 1\arcsec\ (8 kpc) apart, they are likely not, or only
barely, resolved within our $\sim0.75$\arcsec\ seeing. A secondary,
similarly sized but fainter component of B1 lies to the South. We
believe that this region coincides with the Southern extension seen in
\lya\ as well as in faint rest-UV continuum and termed the `blue
filament' in \citet{francis12}. In this paper, we will refer to these
two main optical emission line regions as `B1 North' and `B1 South',
respectively. The regions are surrounded by a much more extended
region of diffuse line emission ($>3\sigma$ per pixel in \oiii\
$\lambda$5007) measuring about $4\arcsec\times5\arcsec$ ($32\times40$
kpc), as shown in the right panel of Fig. \ref{fig:images}.

The spatially resolved \oii, \oiii, and \ha\ emission line maps are
shown in Fig. \ref{fig:maps}. The bright region associated with B1
North has a velocity dispersion of order 800 km s$^{-1}$ (FWHM), seen
both in \oiii\ and \ha\ (middle panels). B1 South is associated with
much smaller line widths of several hundred km s$^{-1}$ (FWHM).  The
velocity shears across B1 are small for all lines, indicating that the
motions are unordered or that we are seeing the source face-on. The
velocities range from $-100$ to $+100$ km s$^{-1}$ (relative to the
redshift at the peak of \oiii\ in B1 North) with some higher and lower
velocity regions mostly in the low S/N pixels on the outskirts. Weak
\oii\ emission is concentrated on B1 North, with a possible extension
to B1 South seen at low S/N.

We extracted integrated spectra from two $1\farcs25\times1\farcs25$
apertures centered on B1 North and South (the apertures are indicated
in the right-hand panel of Fig. \ref{fig:images}), as well as from the
entire B1 region detected in \oiii\ $\lambda$5007 at $>5\sigma$. The
spectra are shown in Fig. \ref{fig:spectra}. The complex line width
profile seen across B1 somewhat complicates the interpretation of the
spectrum integrated over the entire B1 region (top panels) due to the
mixing of the relatively broad lines of B1 North (middle panels) with
the much narrower lines in B1 South (bottom panels). In the remainder,
we will therefore consider the B1 North and South regions
separately. The results of our emission line measurements for B1 North
and South are summarized in Table \ref{tab:lines}.

The \oiii\ lines in B1 North have a FWHM of $724\pm21$ km s$^{-1}$ and
a total \oiii\ flux of $\sim4\times10^{-16}$ erg s$^{-1}$ cm$^{-2}$ at
an average redshift of $z=2.3811$. Due to the relatively large line
widths involved, the line complex around \ha\ (righthand panels in
Fig. \ref{fig:spectra}) is difficult to disentangle. If we force the
line widths of \ha\ and \nii\ to be equal to each other, we find
$816\pm44$ km s$^{-1}$ (FWHM). If we relax this constraint the \ha\
line width is increased by about 100 km s$^{-1}$. An alternative
explanation of the complex \ha\ line profile in B1 North is that
instead of the \nii\ line doublet we are seeing multiple velocity
components of \ha. The two secondary peaks seen redward of the peak in
\ha\ would then have to be offset by about 500 and 1000 km
s$^{-1}$. However, the fact that the line widths of \oiii, \ha, and
\nii\ are all roughly equivalent is good evidence for the presence of
\nii. Also, we do not see any additional velocity components in \oiii,
our strongest line. If we adopt our \ha\ and \nii\ interpretation
(rather than multiple \ha\ peaks), we find a total \ha\ flux of
$3.1\times10^{-16}$ erg s$^{-1}$ cm$^{-2}$ at a redshift consistent
with that of \oiii.

The lines are significantly narrower in B1 South (bottom panels of
Fig. \ref{fig:spectra}). The \oiii\ and \ha\ line widths are
$281\pm15$ and $140\pm34$ km s$^{-1}$, respectively. We do not detect
\nii\ in this region. In neither of the two regions did we detect
\hb. This is consistent with \citet{francis01} who did not detect any
excess flux in the NICMOS narrow-band filter F164N that covers \hb. By
assuming that \hb\ has the same width as \oiii\ and a peak flux twice
the rms measured in the spectral region near \hb, we determine
$2\sigma$ upper limits of $\sim5\times10^{-17}$ and
$\sim2\times10^{-17}$ erg s$^{-1}$ cm$^{-2}$ for B1 North and South,
respectively. Following the same method, we obtained $2\sigma$ upper
limits on the \nii\ line fluxes in B1 South (assuming a line width
similar to \ha).

With these line flux measurements, we can now also directly calculate
the emission line contribution to the broad-band fluxes presented in
\citet{francis01}. We find that the contributions due to \oiii\ and
\ha+\nii\ amount to $\approx$18\% in $H_{160}$ and $\approx32$\% in
$K_S$, respectively, in reasonable agreement with the estimates of
$\sim20$\% of \citet{francis01}.

The observed (i.e. not extinction-corrected) \oiii$\lambda$5007
luminosity of B1 North is $L_{\mathrm{[OIII],obs}}=1.2\times10^{43}$
erg s$^{-1}$, and the total (observed) \oiii\ luminosity of B1 as a
whole is $2.5\times10^{43}$ erg s$^{-1}$. This is at the high end of
the range of \oiii\ luminosities found for LAEs and LBGs at
$z\simeq2-4$
\citep[e.g.][]{teplitz00,pettini01,maschietto08,kuiper11,mclinden11}.
In order to calculate the intrinsic \oiii\ luminosity of B1 North, we
need to correct for dust extinction. We measure a 3$\sigma$ lower
limit on the total extinction based on the Balmer decrement
F(\ha)$/$F(\hb)$\gtrsim$4.13 ($\approx6.2$, 2$\sigma$), implying a
nebular extinction of E(B--V)$\gtrsim$0.34 ($\approx$0.71, 2$\sigma$),
and a extinction-corrected luminosity of
$L_{\mathrm{[OIII],cor}}=3.7\times10^{43}$ erg s$^{-1}$
($\approx$$1.3\times10^{44}$ erg s$^{-1}$, 2$\sigma$).

\subsection{Line ratios and line widths}
\label{sec:shocks}

\begin{figure}[t]
\begin{center}
 \includegraphics[width=\columnwidth]{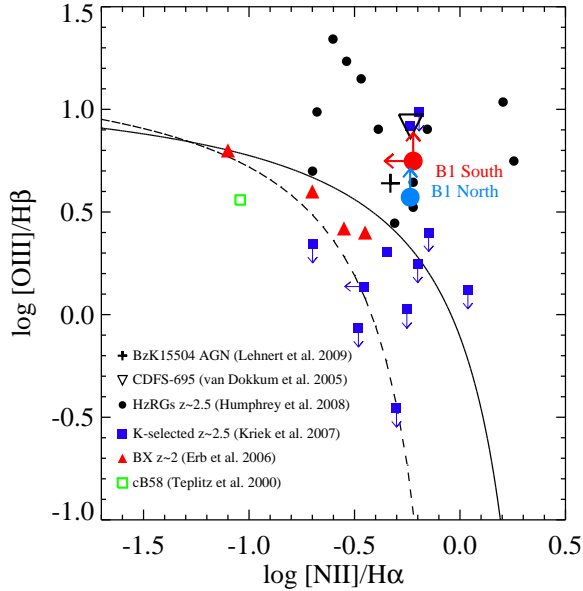}
\end{center}
\caption{\label{fig:bpt}Optical emission line ratio diagram. B1 North
  and South are indicated by the blue and red circles,
  respectively. Other comparison samples shown are the following:
  HzRGs at $z\sim2.5$ (filled circles); BX galaxies at $z\sim2$
  (red triangles); $K$-selected galaxies at $z\sim2.5$ (filled blue
  squares); Lyman break galaxy cB58 (open green triangle); red,
  star-forming galaxy CDFS-695 $z=2.23$ (open upside-down
  triangle); nuclear (AGN) region of BzK15504 at $z=2.38$ (large
  plus). The dashed line marks the boundary between star-forming
  and Seyfert/LINER-like sources from \citet{kauffmann03}. The solid
  line marks the boundary between composite and AGN-like sources from
  \citet{kewley06}.}
\end{figure}

The new measurements of the main optical emission line ratios allow us
to investigate what is the main source of ionization in B1. In
Fig. \ref{fig:bpt} we show the \nii/\ha\ vs. \oiii/\hb\ diagnostics
diagram designed to distinguish between sources in which the main
source of ionizing radiation is the UV emission from hot young stars
and from AGN \citep{baldwin81,veilleux87}. In this diagram,
star-forming galaxies tend to lie to the left of the dashed line which
separates sources with and without Seyfert- or LINER-like
characteristics \citep{kauffmann03}. The solid line further aims to
separate galaxies of mixed or composite spectral type from those that
are pure AGN-like \citep{kewley06}. The values obtained for B1 North
and South are indicated by the blue and red circles. For
non-detections of \hb\ we have used the $3\sigma$ upper limit for B1
North, and the intrinsic Balmer ratio of \ha/\hb$=$2.8 for B1
South. For B1 South we furthermore used the $3\sigma$ upper limit on
\nii. Both regions of B1 lie in the AGN-dominated region of the
diagram.

Unfortunately, the double limits on B1 South prohibit us of making
strong conclusions about this region, due to the non-detections of
both \hb\ and \nii. Although the line ratios of B1 South are
consistent with those of B1 North, the dominant source of ionization
could, besides an AGN, be a starburst given that we only have an upper
limit on the \nii/\ha\ line ratio. The high \oiii/\hb\ then implies
that it is a very low metallicity starburst
\citep[see][]{kewley01}. We consider this somewhat unlikely, however,
given that the \nii/\ha\ measured from integrated spectra of
star-forming galaxies are typically higher than those seen in pure
\hii\ regions due to ionized material in a diffuse component
\citep{lehnert94}, and B1 South covers a significant region of diffuse
emission. In our discussion below, we will focus our attention on B1
North (the region with the strongest line emission and the largest
line widths), but we note that our results would not change
significantly if instead we had performed our measurements over both
regions or integrated over the entire region of B1.

The relatively large values found for the line ratios in B1 North are
inconsistent with stellar photoionization, and indicative of
photoionization by an AGN. B1 North has line ratios that are very
similar to, for example, the massive AGN-hosting starburst galaxy from
\citet{lehnert09}, or the population of $z\sim2.5$ radio galaxies from
\citet{humphrey08}. The AGN-interpretation is confirmed by the rather
weak \oii\ emission. The (3$\sigma$) lower limit on the
extinction-corrected \oii\ luminosity is $\sim1\times10^{43}$ erg
s$^{-1}$, implying a extinction-corrected \oiii/\oii\ ratio of
$\approx$3.8. This line ratio is a sensitive diagnostic of the
ionization parameter, and efficiently separates Seyfert-like objects
from star-forming galaxies and LINERS at low redshift
\citep{kewley06}. Although we do not have any constraints on \oi/\ha\
that could further separate the AGN from star-forming and composite
galaxies, at an \oiii/\oii$>$1 the main ambiguity is between Seyferts
and the rather rare class of very low metallicity starbursts. While
such low metallicity starbursts (and corresponding high \oiii/\oii\
ratios) are common among the population of LBGs and LAEs at high
redshift \citep{nakajima12}, B1 is a large, dusty, and
high-metallicity source that is very different from such high redshift
galaxies.

In principle, shocks offer an alternative possibility for the line
ratios in B1 North. Shocks would also be consistent with the broad
line widths of $\sim$800 km s$^{-1}$ observed. However, the
discrimination between shocks and AGN photoionization is a
longstanding problem in astrophysics, even when multiple UV and
optical emission lines are available. With just the three line ratios
detected in B1, a quantitative distinction is just not
possible. Comparing the extinction-corrected line ratios to photo- and
shock-ionization model predictions presented in \citet{humphrey08}, we
conclude that B1 North lies in the region of line ratios where power
law photoionization and shock-precursor models overlap. This means
that photoionization (either from the shock precursor or from the AGN)
rather than collisional excitation due to the shocks dominates the
excitation of the gas.

\begin{figure}[t]
\begin{center}
\includegraphics[width=\columnwidth]{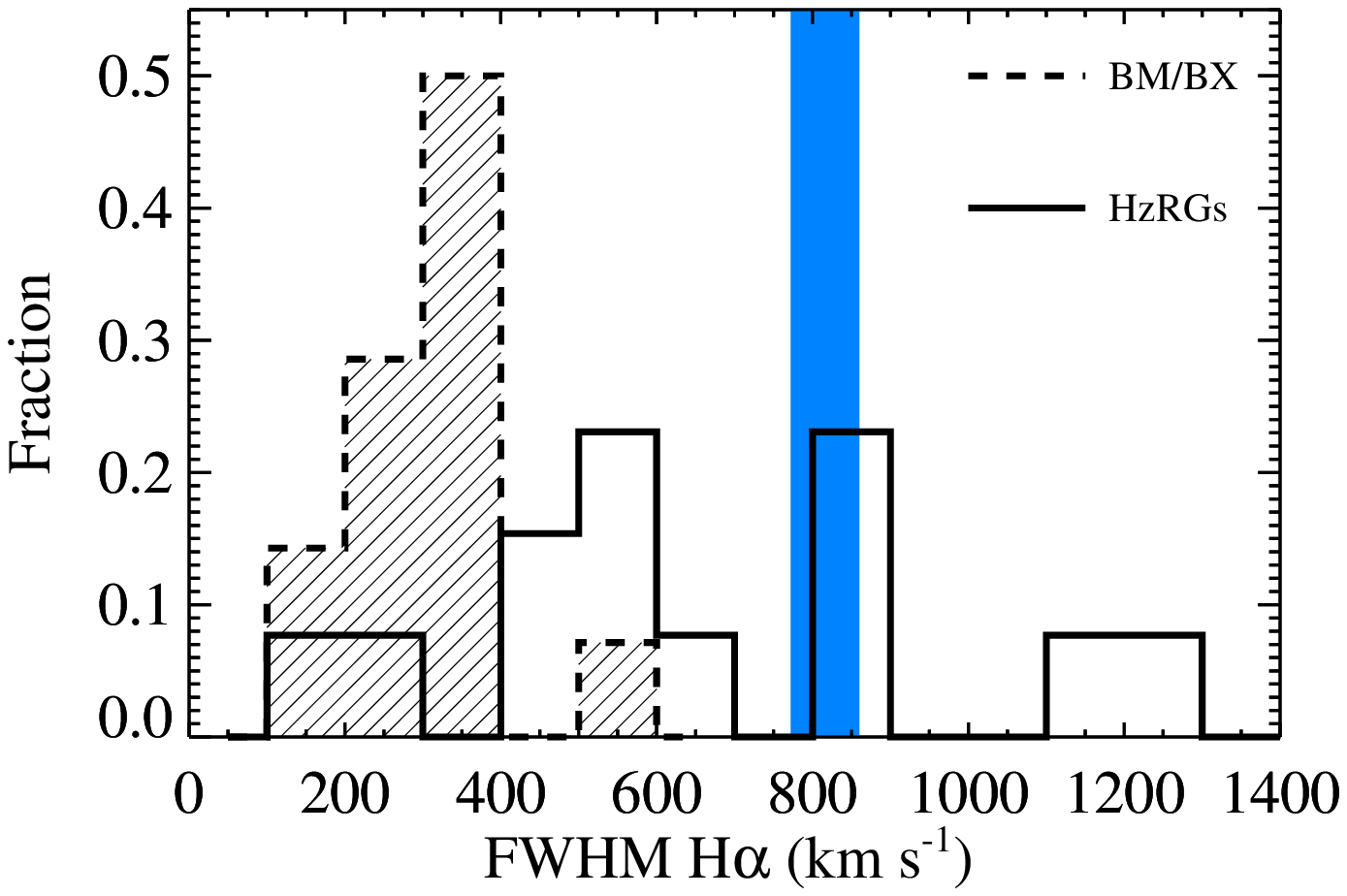}\\
\includegraphics[width=\columnwidth]{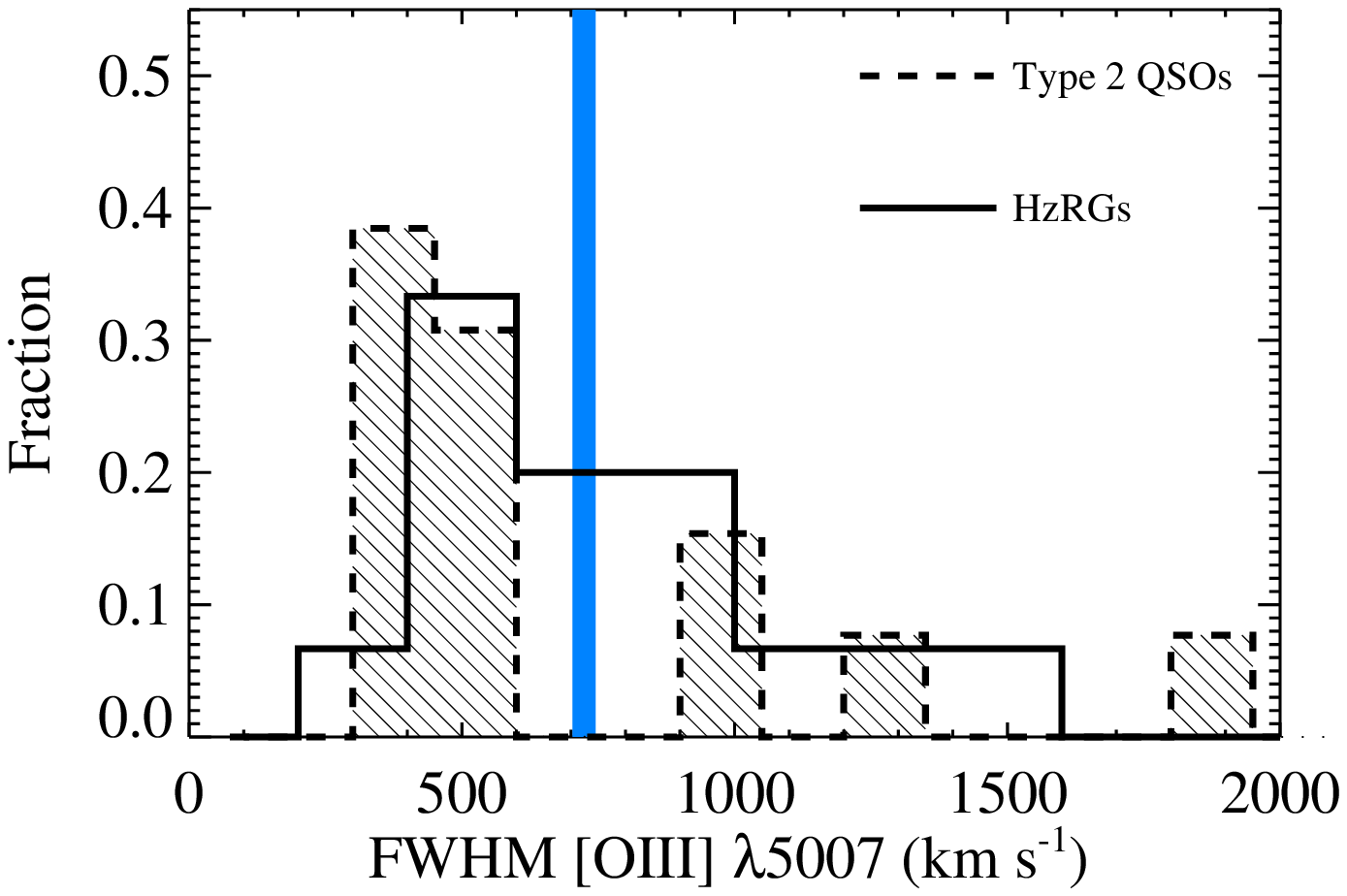}
\end{center}
\caption{\label{fig:fwhm}The line velocity widths. Top and bottom
  panels show the FWHMs of \ha\ and \oiii\ measured in B1 North (blue
  shaded region). For comparison, in the top panel we show the \ha\
  FWHM distributions for BM/BX galaxies at $z\sim2$ \citep[hatched
  histogram,][]{forster06} and for radio galaxies at $z\sim2.5$
  \citep[open histogram,][]{humphrey08}. In the bottom panel, we
  compare with the \oiii\ FWHM distributions for Type 2 QSOs at
  $z=0.3-0.6$ \citep[hatched histogram,][]{villar-martin11} and for
  radio galaxies at $z\sim2.5$ \citep[open histogram,][]{humphrey08}. The large gas motions of B1 North are
  most consistent with those observed in high redshift radio galaxies
  and Type 2 quasars, and not with star-forming galaxies.}
\end{figure}

In Fig. \ref{fig:fwhm} we compare the widths of \ha\ and \oiii\ with
those of star-forming galaxies and AGN. For \ha, we compare with the
distribution of FWHM as measured for $z\sim2$ star-forming galaxies by
\citet{forster06}, and for $z\sim2.5$ HzRGs by \citet{humphrey08} (top
panel). The line widths of B1 North are much larger than those found
for typical $z\sim2$ galaxies, and more comparable to those of
HzRGs. The \oiii\ width can be compared to the distributions measured
for the sample of Type 2 quasars at $z\sim0.3-0.6$ studied by
\citet{villar-martin11} and the HzRGs from \citet{humphrey08} (bottom
panel). The kinematics of B1 North are again similar to the mean line
widths observed for HzRGs. It is interesting to note that the narrow
lines of Type II QSOs reach similar line widths as those of HzRGs,
implying that large turbulent motions exist even in sources without
strong radio jets. Could the large velocity dispersions and large line
ratios be explained by shocks arising from a starburst superwind? This
is highly unlikely given that the ratio of mechanical to bolometric
energy output from a continuously star-forming galaxy is small on a
dynamical time scale \citep{leitherer99,letiran11}, and thus we would
expect to see the \hii\ region-like emission on some scale in B1,
which is not seen.

As we will show below, the simplest scenario that explains most or all
of the properties of B1 is photoionization by a luminous obscured
quasar. 

\begin{figure}[t]
\begin{center}
\includegraphics[width=\columnwidth]{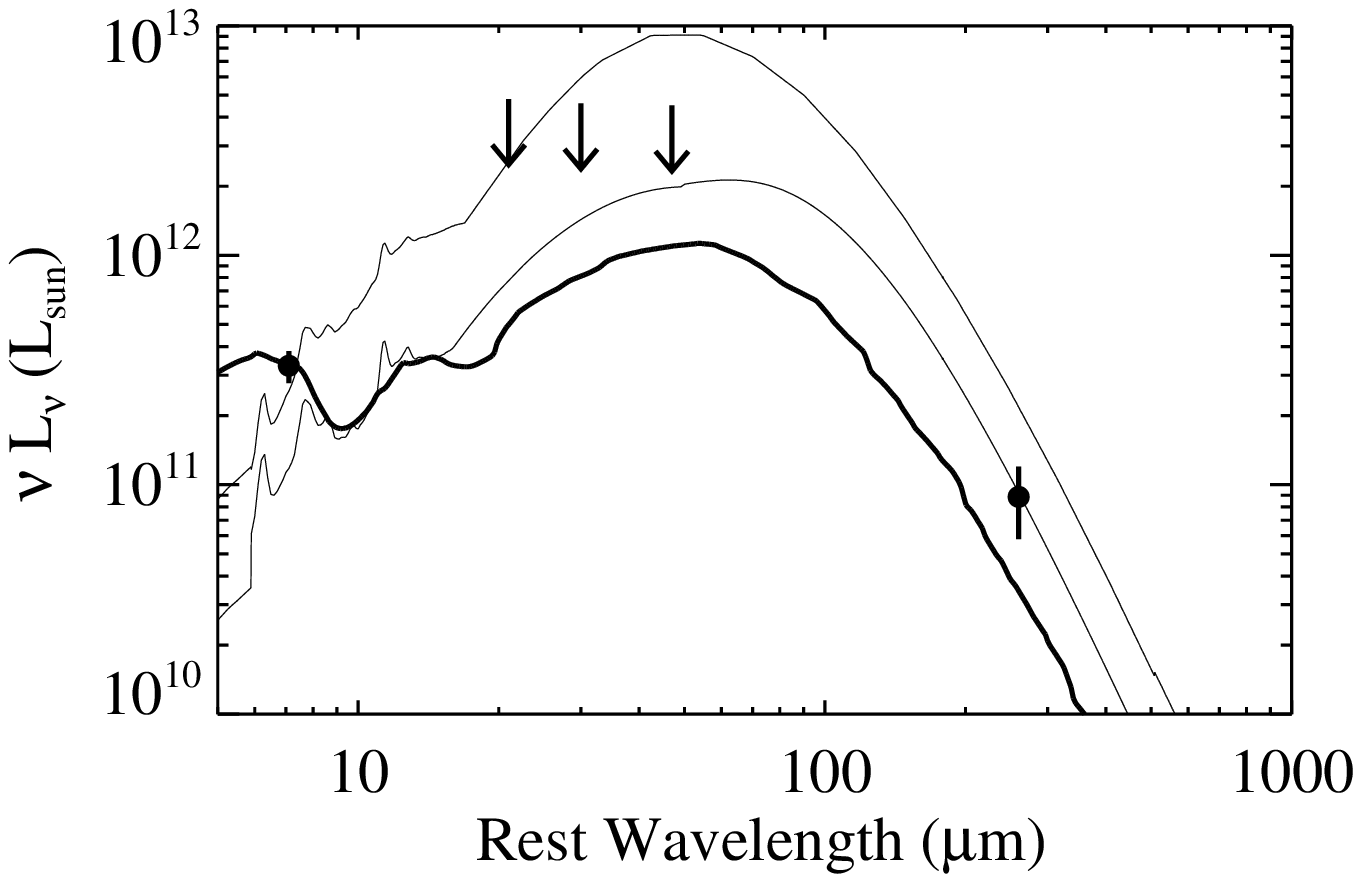}
\end{center}
\caption{\label{fig:ir}Constraints on the infrared energy distribution
of B1. Data-points are the detections at observed 24$\mu$m
\citep{colbert11} and 870 $\mu$m \citep{beelen08}. The upper limits
(5$\sigma$) based on the Herschel/PACS reductions presented in this paper are
shown as downward arrows. Upper and lower thin lines indicate the \citet{chary01}
models of $L_{8-1000\mu m}=1.3\times10^{13}$ and $3.6\times10^{12}$
$L_\odot$ that best match either the 24$\mu$m or 870$\mu$m data-points,
respectively. The thick line shows the spectral energy distribution of
the QSO-like local ULIRG Mrk 231, scaled to the 24$\mu$m luminosity of B1.}
\end{figure}

\subsection{Evidence for a luminous, obscured AGN}
\label{sec:agn}

The \civ\ emission, the AGN-like line ratios, and the relatively high
(compared to typical starburst galaxies) velocity dispersions suggest
that B1 North hosts a luminous AGN. Let us consider the situation in
which the dominant source of ionization is indeed a hidden
quasar. Because the BPT line ratios of B1 are consistent with
photo-ionization by an AGN, in this scenario we can safely ignore the
contribution from strong ionizing radiation fields due to star
formation, and thus use the \oiii\ luminosity as a direct proxy for
the bolometric luminosity of the quasar. \citet{heckman04} find that
$L_{\mathrm{bol,agn}}\approx3500L_{\mathrm{[OIII]}}$ with a scatter of
0.38 dex. Using the observed (i.e. not extinction-corrected)
luminosity of B1 North we find
$L_{\mathrm{bol}}\approx4.2\times10^{46}$ erg s$^{-1}$. Alternatively,
\citet{lamastra09} derived that
$L_{\mathrm{bol,agn}}\approx454L_{\mathrm{[OIII],cor}}$, where
$L_{\mathrm{[OIII],cor}}$ is the luminosity corrected for extinction
based on the Balmer decrement and the constant was determined for AGN
having a extinction-corrected \oiii\ luminosity of 10$^{42-44}$ erg
s$^{-1}$. This gives $L_{\mathrm{bol}}\approx1.7\times10^{46}$ erg
s$^{-1}$. Both estimates would thus imply the presence of a luminous
quasar \citep[e.g.][]{lamastra09,harrison12,kim13}.

The source has a significant rest-frame 8 $\mu$m luminosity of
$\sim3.3\times10^{11}$ $L_\odot$ \citep{colbert06}, previously
attributed to dust heated by a possible obscured AGN \citep{colbert06}
(a conclusion later disfavored by \citet{francis12}). Scaling from the
8 $\mu$m luminosity gives $L_{IR}\simeq1-3\times10^{12}$ $L_\odot$
assuming the source is an AGN or starburst-AGN composite
\citep{wu10}. An obscured QSO is also consistent with the limits at
longer infrared wavelengths, assuming that our hidden quasar has a
spectral energy distribution similar to, e.g., Mrk 231, a local ULIRG
with a QSO-like spectrum (see Fig. \ref{fig:ir}).

B1 has not been detected in the X-rays. \citet{francis04b} placed a
limit on the hard X-ray flux of $<10^{-15}$ erg cm$^{-2}$ s$^{-1}$,
corresponding to a rest-frame luminosity of
$L_{2-24keV}<4.3\times10^{43}$ erg s$^{-1}$ assuming no absorption. A
quasar having a bolometric luminosity of a few times $10^{46}$ erg
s$^{-1}$ should have an X-ray luminosity of $\sim10^{45}$ erg $s^{-1}$
\citep{lamastra09}, at least 23$\times$ higher than that observed. The
non-detection in the X-ray suggests that the quasar is optically-thick
(or even Compton-thick, if the intrinsic X-ray luminosity is
$\sim60\times$\ higher than the upper limit).

\subsection{Limits on star formation}
\label{sec:sfr}

What is the role of star formation in the B1 emission line nebula? A
relatively unbiased SFR measurement can perhaps be made best in the
far-infrared (FIR), regardless of whether an AGN is present or not
\citep{alexander05,menendez-delmestre09}.  As mentioned, the source
has a significant rest-frame 8 $\mu$m luminosity of
$\sim3.3\times10^{11}$ $L_\odot$ \citep{colbert06}, implying
$L_{IR}\simeq1-3\times10^{12}$ $L_\odot$ \citep{wu10}.  The only
longer wavelength constraint available in the literature is a flux
density of 2.3 mJy at 870 $\mu$m obtained for a stack of several LABs
\citep{beelen08}, implying $\nu L_\nu\sim8.8\times10^{10}$ $L_\odot$
at a rest-frame of about 260 $\mu$m. In addition, we have used
archival Herschel/PACS data of B1, finding that it is not detected in
any of the PACS channels. We derive upper limits (5$\sigma$) of 9.9,
13.6, and 21.3 mJy at 70, 100, and 160 $\mu$m, respectively. As shown
in Fig. \ref{fig:ir}, we can thus rule out IR luminosities of
$\gtrsim4\times10^{12}$ $L_\odot$, based on the PACS limits as well as
the 870 $\mu$m constraint. In Fig. \ref{fig:ir} we have indicated the
spectral energy distribution of Mrk 231, a local ULIRG with a QSO-like
spectrum, scaled to the rest 8$\mu$m luminosity of B1 (thick solid
line). While such an obscured QSO spectrum is consistent with the PACS
limits, B1 may require an additional contribution from cold dust
heated by star formation to explain the rest 260 $\mu$m
emission. Ignoring the QSO contribution at this wavelength (expected
to be $\sim3\times$ lower), and interpreting the cold dust detection
as a measure of star formation, we estimate a SFR of $\sim$570
$M_\odot$ yr$^{-1}$ for a Salpeter IMF (380 $M_\odot$ yr$^{-1}$ for a
Kroupa IMF). Because this ignores any AGN contribution, we consider
this an upper limit, and note that B1 is consistent with having no
star formation at all.

\section{Discussion}
\label{sec:lya}

\subsection{B1 as an AGN-powered \lya\ blob}

The scenario in which B1 hosts a luminous AGN is consistent with most
of its observed properties. Its overall spectral characteristics are,
in fact, very similar to those of other luminous, obscured quasars at
high redshift \citep[e.g.][]{nesvadba11,harrison12,kim13}. The \oiii\
luminosity is comparable to that of the powerful obscured quasars
SWIRE J022513.90--043419.9 (SW022513) and SWIRE J022550.67--042142
(SW022550) at $z\sim3.5$ studied by \citet{nesvadba11} and
\citet{polletta11}. Both QSOs are Compton-thick
\citep{polletta08}. The SWIRE QSOs have narrow line region \oiii\ line
widths of order 1000 km s$^{-1}$ (FWHM), and the \oiii\ emission in SW022513 extends out to at least 10 kpc, 
consistent with that of B1 North. The line ratios, and the \oiii\ size
and luminosity are also similar to those of the extended emission line
regions observed around low redshift quasars
\citep[e.g.][]{fu09,villar-martin11}. In fact, narrow line regions of around 10 kpc are typical of
luminous obscured quasars, while regions of up to several tens of kpc
are characteristic for the most luminous high redshift QSOs with
strong \oiii\ emission \citep{netzer04}. The narrow line regions are
also often asymmetric \citep[e.g.][]{lehnert09,nesvadba11}, which may
come as no surprise given that most likely there are strong gradients
in the local ionizing radiation field due to the high extinction by
dust.

Now that we have found evidence for an extremely luminous hidden
quasar we can finally attempt to estimate the number of ionizing
photons available to power the observed \lya\ nebulosity of
$8\times10^{43}$ erg s$^{-1}$ in the scenario in which the hidden
quasar is the main source of ionizing power. In Sect. \ref{sec:agn} we
estimated that $L_{bol,agn}\simeq1.7-4.2\times10^{46}$ erg s$^{-1}$.
We calculate the ionizing luminosity from 200--912\AA\ by assuming a
typical radio-quiet quasar spectrum of the form
$f_\nu\propto\nu^\alpha$ with $\alpha=-0.5$ for log($\nu$) in the
range 9--15 and $\alpha=-1.5$ for log($\nu$) in the range 15--19
\citep{richards06}. The total ionizing luminosity is then
$\sim4\times10^{45}$ erg s$^{-1}$, sufficient to provide a luminosity
of $L_{Ly\alpha}\approx3\times10^{45}$ erg $s^{-1}$. Here we have
assumed that the fraction of ionizing photons that will cascade to
\lya\ photons is 68\% (case B recombination). Although this number
exceeds the actual amount of ionizing radiation likely to be available
due to the absorption by dust by a factor of $\sim10$, our hidden
quasar scenario still appears to be more than sufficient to power the
nebula even without the aid of additional ionizing photons from, e.g.,
star formation.

Galaxies that have SFRs large enough to provide sufficient ionizing
photons to power the \lya\ luminosities observed in LABs are typically
heavily obscured SMGs. In these galaxies, the large amounts of dust
will prohibit significant escape of \lya\ photons, although \lya\ has
been detected in a number of cases, presumably due to strong spatial
variations in the dust coverage \citep{chapman03,chapman04}. Powerful
obscured quasars such as the one we discovered in B1, offer an
additional ionizing radiation field that is two orders of magnitude
larger than that required by the observed \lya. This means that, even
for the same escape fraction, they can power luminous LABs despite
them often being dusty objects as well.

\begin{table*}[t]
\begin{center}
\caption[]{\label{tab:census}A census of luminous extended \lya\
  emission in LABs, Radio Galaxies, and Quasars}
\begin{tabular}{lccrll}
\hline
\hline
ID                     & Redshift  & log$L_{Ly\alpha}$ & size$^a$ &Notes$^b$ & References$^c$ \\
                       &           & (erg s$^{-1}$) & (kpc) & & \\
\hline
\hline
\multicolumn{6}{c}{\lya\ blobs}\\
\hline
\hline
SSA22--Sb1--LAB1       & 3.10  & 44.0 & 175   & (Type II AGN; detected in X-ray stack) & G09\\
SSA22--Sb3--LAB1       & 3.10  & 44.3 & 126   & RQ-QSO; broad lines & S07, M11\\
LAB1709+5913           & 2.83  & 44.3 & 95    & --$^d$ & Sm07, Sm08\\
SST24J1434110+331733 & 2.66 & 44.2 & 160 & Type II AGN; narrow \civ, \heii, power-law SED & D05\\
AMS05                & 2.85 & 44.2 & 80  & Type II AGN; strong 24 $\mu$m & Sm09\\
LAB1\_J2143--4423 (B1)  & 2.38 & 43.9 & 137   & Type II AGN; narrow \civ, BPT & F96, C06, C11, This paper\\
CDFS-LAB01           & 2.3  & 43.9 & 60 & (Type II AGN; narrow \civ, \heii) & Y10, Y11\\
LAB6\_J2143--4423 (B6)  & 2.38 & 43.8 & 64    & Type II AGN; narrow \heii, power-law SED  & Sc09,C11\\
LAB5\_J2143--4423 (B5)  & 2.38 & 43.8 & 56    &     --$^d$          & C11\\  
SSA22--Sb1--LAB2       & 3.09  & 43.8 & 157   & Type II AGN; X-ray & BS04\\
SSA22--Sb6--LAB1       & 3.10  & 43.8 & 166   &     --$^d$      & M11\\  
SSA22--Sb1--LAB3       & 3.10  & 43.7 & 103   & Type II AGN; X-ray & G09\\
GOODS-N--LAB1         & 3.08  & 43.7 & 124   & RQ-QSO; broad lines, X-ray & B02, M11\\
53W002-Object 18      & 2.39 & 43.7 & $>$40    & Type II AGN; narrow \nv, \civ, \heii & P96, K99\\  
Yang--LAB3            & 2.32 & 43.7 & 61     & Type I AGN; broad \civ, X-ray & Y09, G09\\  
PRG1                 & 1.67 & 43.7  & 56 & (Type II AGN; narrow \civ, \heii, \ciii) & P09\\
\hline
\hline
\multicolumn{6}{c}{Radio Galaxies$^e$}\\
\hline
\hline
MRC 1138--262  & 2.16 &  45.4 & 250  &  Radio Galaxy & V07\\
4C41.17 & 3.80 & 45.2 & 190 &  Radio Galaxy & R03\\
4C60.07 & 3.79 & 45.1 & 68 &  Radio Galaxy & R03\\
BRL 1602--174  & 2.04 &  44.9 & 90  & Radio Galaxy & V07\\
B2 0902+34 & 3.39 & 44.8 & 80 &  Radio Galaxy & R03\\
TN J1338--1942 & 4.11 & 44.7 & 130  &  Radio Galaxy  & V07\\
3C 294         & 1.79 & 44.5  & 170 & Radio Galaxy & M90\\
TN J2009--3040 & 3.16 & 44.5 & 40  &  Radio Galaxy & V07\\
MRC 1558-–003 & 2.52 & 44.5 & 84 &  Radio Galaxy & VM07\\
MRC 0943--242  & 2.92 & 44.4 & 50  &  Radio Galaxy & V07\\
MRC 2025–-218 & 2.63 & 44.2 & 55 & Radio Galaxy & VM07\\
MRC 0052--241  & 2.86 &  43.9 & 35  & Radio Galaxy & V07\\
MRC 2048--272  & 2.06 &  43.8 & 70  &  Radio Galaxy & V07\\
MRC 0316--257  & 3.13 & 43.8 & 35  & Radio Galaxy & V07\\
\hline
\hline
\multicolumn{6}{c}{Quasars}\\
\hline
\hline
Heckman et al. Sample & $\langle2.2\rangle$ & $\langle43.7\rangle$&$\langle97\rangle$ & RL-QSO & H91\\
SDSS J21474--0838 & 4.59  & 44.3 & 51 & RQ-QSO & N12\\
Q1425+606         & 3.204 & 43.9 & 34 & RQ-QSO & C06\\
Q1759+7539        & 3.049 & 43.9 & 60 & RL-QSO & C06\\
\hline
\hline
\end{tabular}
\end{center}
\begin{scriptsize}
$^a$ Major axis size.\\
$^b$ Notes are as follows. Type II AGN: evidence of an
obscured AGN, Type I AGN: evidence of an unobscured AGN, --: no
evidence of AGN, QSO: optically-or radio selected
QSO (RL-QSO: Radio-loud QSO, RQ-QSO: Radio-quiet QSO). For objects for which the identification is unclear the type  
is given in parentheses.\\
$^c$ References. F96: \citet{francis96}, C06: \citet{colbert06}, C11:
\citet{colbert11}, Sc09: \citet{scarlata09}, G09: \citet{geach09},
BS04: \citet{basu-zych04}, M11: \citet{matsuda11}, S07:
\citet{shen07}, D05: \citet{dey05}, Sm07: \citet{smith07}, Sm08: \citet{smith08}, Sm09: \citet{smith09}, P96:
\citet{pascarelle96}, K99: \citet{keel99}, Y09: \citet{yang09}, Y10:
\citet{yang10}, Y11: \citet{yang11}, P09: \citet{prescott09}, V07:
\citet{venemans07}, VM07: \citet{villar-martin07b}, R03:
\citet{reuland03}, M90: \citet{mccarthy90}, Chr06:
\citet{christensen06}, N12: \citet{north12}, H91: \citet{heckman91}.\\
$^d$ No classification given either indicates no
evidence of an AGN or no AGN diagnostics available.\\
$^e$ It is important to note that this subset of well-studied HzRGs
occupies the high end of the ranges in \lya\ luminosity and major axis
size of the general population of high redshift radio galaxies.\\ 
\end{scriptsize}
\end{table*}

\subsection{Luminous \lya\ blobs harbor luminous Type II AGN}

We have shown that (i) B1 harbors an extremely luminous obscured
(i.e. Type II) quasar, (ii) only a few percent of the bolometric
luminosity is sufficient to power the observed \lya\ line luminosity
of $\sim10^{44}$ erg s$^{-1}$, and (iii) the extended narrow line
region is not unlike that observed toward other quasars at high
redshift. The significance of this new finding becomes particularly
clear when we perform a census of other \lya\ blobs (LABs) found in
the literature. Here, we will limit ourselves to the class of truly
powerful objects having $L_{Ly\alpha}\gtrsim5\times10^{43}$ erg
s$^{-1}$ and sizes of order 50-100 kpc in order to avoid the far more
common population of lower luminosity and smaller \lya\ emitters for
which the energy sources become much more ambiguous. The results of
our census are summarized in Table \ref{tab:census}, with details on
the selected targets as follows:

\smallskip
\noindent
1. Several LABs were found as part of the same J2142--4423
``proto-cluster'' that includes B1 studied in this paper. Strong
evidence for B1 as a powerful obscured AGN is presented in this
paper. The second brightest object (B6) hosts a Type II AGN as well
\citep{scarlata09}. There is no evidence to date that B5 contains an
AGN \citep{colbert11}.  

\smallskip
\noindent
2. There are five extended LABs in the SSA22 protocluster at $z=3.1$
\citep{steidel00} that fulfill our criteria (SSA22--Sb1--LAB1,
SSA22--Sb1--LAB2, SSA22--Sb1--LAB3 using the nomenclature of
\citet{matsuda11}. SSA22--Sb1--LAB2 and SSA22--Sb1--LAB3 have been
identified as obscured AGN \citep{basu-zych04,geach09,webb09}. 
SSA22--Sb1--LAB1 may host a heavily obscured AGN as well based on the
detection in a stack of X-ray images of several undetected blobs
\citep{geach09}, but we list it as ambiguous in Table \ref{tab:census}. 

\smallskip
\noindent
3. \citet{matsuda11} performed a large-area narrow-band search
targeting several fields. By expanding the area around the
above-mentioned SSA22 structure, they found two more luminous extended
LABs: SSA22--Sb3--LAB1 is identified with a luminous optical
QSO. SSA22--Sb6--LAB1 is not currently known to harbor any
AGN. Another object, GOODS-N-LAB1, was found to be associated with a
previously known optical QSO. Because the two LABs associated with
QSOs were discovered serendipitously from the narrow-band survey of
\citet{matsuda11}, we have listed these objects here rather than in
the QSO section of Table \ref{tab:census}.

\smallskip
\noindent
4. \citet{dey05} found a luminous, extended nebula at $z=2.7$
(SST24J1434110+331733), most likely harboring a dust-enshrouded quasar
based on an analysis of the optical-IR spectral energy distribution.

\smallskip
\noindent
5. \citet{smith07} found a $z=2.83$ LAB in the Spitzer First
Look Survey. LAB1709+5913 displays no evidence for either AGN or
significant star formation that could explain its luminous \lya\
nebulosity. Instead, the \lya\ could originate from the heating and
cooling of gas in an accretion flow. \citet{smith09} studied a
luminous, extended LAB (AMS05) associated with a Type II QSO at
$z=2.85$ in the {\it Spitzer} First Look Survey. This LAB is
particularly interesting given its resemblance to the \lya\ halos
associated with HzRGs despite the fact that the AGN is orders of
magnitude fainter in the radio compared to the FRII sources associated
with HzRGs. Similar to B1, the QSO bolometric luminosity is two orders
of magnitude higher than that emitted by the \lya\ photons. If AMS05
would lie at $z=2.38$ its radio flux at 1.4 GHz would be below the
upper limit on the flux of B1 \citep[3.3 mJy, see][]{francis96}.

\smallskip
\noindent
6. \citet{pascarelle96} and \citet{keel99} have studied a luminous,
extended nebula at $z=2.4$ (53W002-Object 18) showing broad and narrow lines
from an AGN. \citet{keel02} detected spatially extended \oiii\
emission with a luminosity that is comparable to that of B1. 

\smallskip
\noindent
7. \citet{yang09} studied four $z=2.3$ LABs in the Bootes field, one
of which (Yang--LAB3) is luminous enough to fulfill our criteria. This
LAB has broad UV emission lines and an X-ray detection indicative of a
luminous unobscured AGN \citep[see also][]{geach09}. \citet{yang10,yang11}
studied CDFS-LAB01 at $z\approx2.3$, detecting narrow \civ\ and
\heii\ lines. We tentatively classify this source as a Type II AGN, but
list it as ambiguous in Table \ref{tab:census}.    

\smallskip
\noindent
8. PRG1 is a $z=1.67$ LAB showing extended \heii\ and weak \civ\ and
\ciii\ emission lines \citep{prescott09}. The source has not been
detected in the X-ray, but the limits are not very strong for typical
Seyfert galaxies. The source was not detected at 24$\mu$m or in the
radio. This could be reconciled with the AGN-like FUV line ratios if
it hosts a heavily obscured AGN. We tentatively classify this source as a Type II AGN, but
list it as ambiguous in Table \ref{tab:census}.  

\smallskip
\noindent
9. Luminous, extended \lya\ nebulae have also been found around
quasars. In Table \ref{tab:census} we provide the median values
measured for the sample of radio-loud QSOs from \citet{heckman91}, as
well as a number of radio-loud and radio-quiet QSOs studied by other
groups \citep[e.g.][]{christensen06,north12}.

\smallskip
\noindent
10. Also in Table \ref{tab:census} we list a representative sample of
high redshift radio galaxies and properties of their \lya\ nebulae
selected from the literature
\citep[e.g.][]{mccarthy90,venemans07,villar-martin07b,reuland03}. As
already stated in \S\ref{sec:intro} there is overwhelming evidence
that the main source of ionization in this class of objects is the
photoionization by an obscured AGN. Although HzRGs show a number of
extreme phenomena that are typically not seen in their radio-quiet
counterparts, these phenomena are almost exclusively found in the
close vicinity of the radio jets.

\medskip The census of narrow-band selected LABs presented in the top
part of Table \ref{tab:census} can be considered a complete sample for
luminosities of $\gtrsim5\times10^{43}$ erg s$^{-1}$. It is clear that
at this high luminosity threshold, the presence of luminous AGN is a
common theme among the LABs (10/16 confirmed AGN, 13/16 when including
tentative AGN). The luminosities and sizes of the LABs are comparable
to those of HzRGs, eventhough the brightest and largest \lya\ nebulae
are always associated with HzRGs. They are also similar to those found
near some radio-quiet and radio-loud quasars. While the majority of
radio-loud quasars appear to show extended \lya\ nebulae
\citep[e.g.][]{heckman91} with properties that are, on average, very
similar to those of the LABs, it is not yet clear whether the same is
true for their radio-quiet counterparts due to the relatively small
sample sizes \citep{christensen06,north12}. The detailed comparison
between the QSOs and other classes of sources with extended \lya\
nebulae (e.g. HzRGs and LABs) is particularly difficult because of
sample selection, geometric effects, obscuration, and because in QSOs
the line-of-sight is dominated by luminous \lya\ emission from the AGN
broad line region.  Therefore, the QSOs listed in the third part of
Table \ref{tab:census} do not form a complete sample, but they were
specifically chosen to illustrate that the basic properties of
extended \lya\ nebulae are comparable among the LABs, HzRGs, and (at
least some) QSOs.

If we look at the evidence presented in Table \ref{tab:census}, at
least 63\% (81\% when including tentative AGN) of luminous LABs are
associated with luminous AGN. These numbers are lower limits because
not all LABs have been observed at similar depths or wavelengths.  The
AGN fraction is thus very high, especially since the AGN duty cycles
(the time spent by a black hole in the `active' state) are believed to
be relatively short (10-100 Myr). This must mean that \lya\ surveys
always tend to find those objects that are in the active state, and
therefore are providing a plentiful supply of ionizing photons to
their surrounding gaseous medium. Also, because the AGN duty cycles
are short (compared to a galaxy building time scale), the LABs that
are visible at any given epoch probably represent a larger population
of gaseous halos that are not being actively illuminated, at least not
to an extent that would give rise to similar luminosities and sizes as
those of the LABs listed in Table \ref{tab:census}.

Both the HzRGs and the radio-quiet LABs have been associated with
overdense regions in the high redshift large-scale structure
\citep{francis01,francis12,pentericci99,pentericci00,steidel00,venemans02,venemans07,matsuda04,matsuda09,overzier06,overzier08,saito06,miley04,miley06,prescott08,prescott12,erb11},
leading to a popular hypothesis that the LAB phenomenon is perhaps
linked to \lya\ cooling radiation in massive dark matter halos as
predicted by theory. However, our census of the most luminous LABs
presented above appears to indicate that the primary correlation is
that between (luminous) LABs and AGN. \citet{geach09} reached a very
similar conclusion based on deep X-ray observations of the SSA22 field
targeting a great number of extended LAEs discovered by
\citet{matsuda04}. They derived an AGN fraction of at least 17\%, but
this is considered a strict lower limit as the sample is dominated by
sources down to $L_{Ly\alpha}\simeq6\times10^{42}$ erg s$^{-1}$ that
are X-ray faint. In fact, if indirect evidence for AGN based on IR
emission is included, the AGN fraction within the SSA22 field is
$\sim$30\%, and if a higher luminosity cut is placed on \lya\ to match
the selection of \citet{yang09}, the AGN fraction is $\sim$50\%. As we
have shown, if we increase our \lya\ luminosity threshold even
further, the AGN fraction among all luminous LABs found in the
literature approaches 100\%. This is consistent with the recent study
of \citet{bridge12}, who studied a population of rare, dusty objects
having high IR luminosities and warm IR colors suggestive of intense
AGN activity \citep[see also][]{wu12}. 90\% show \lya\ in emission, with 37\% showing \lya\
(luminosities $10^{42-44}$ erg s$^{-1}$) extended over $>$30 kpc,
i.e. comparable to the LAB selection criteria of
\citet{matsuda04,matsuda11}.

If, as we suggest, the primary correlation is between LABs and AGN,
then the correlation between LABs and environment could be a secondary
one between environment and AGN. The fact that excesses of LABs are
primarily found in overdense regions would then simply be due to the
fact that overdense regions are overdense in massive galaxies that are
more frequently associated with AGN \citep[see also][]{matsuda11}. The
extra source of ionizing photons provided by these AGN will further
boost the correlation for the most luminous LABs, even if some LABs
can be powered by ionizing radiation from star-formation alone. A
correlation between LABs and environment is also expected in the cold
streams models of \citet{dijkstra09}. In these models, the most
luminous \lya\ halos are always associated with cooling radiation from
the most massive halos, for which other simulations typically predict
that they are host to massive galaxies (and possibly AGN). There is
one important difference though. The cosmological cold streams have a
duty cycle of $\approx$1, much longer than that of the AGN. Therefore,
if the LABs were powered by gravitational heating (and subsequent
cooling), then we would expect a (significant) fraction of the LABs to
be associated with massive halos in which the black hole is
inactive. The fact that we find an extremely high fraction of AGN,
strongly suggests that the AGN is the main driver of the \lya\
luminosity in these LABs. Even in the cases where no AGN or extensive
star formation is observed, \lya\ cooling radiation is not the only
alternative explanation. In the nearby universe, examples of massive
extended clouds with AGN-photoionized line-ratios have been found
around galaxies in which the AGN has apparently switched off on short
time-scales \citep{keel12}, illustrating that episodic AGN activity
may need to be considered as well when interpreting high redshift
LABs.

Irrespective of what is powering the \lya\ emission, the photons must
be coming from relative cold gas ($10^{4-5}$ K). Given that the LABs
are probably associated with massive halos with $T_{vir}>10^6$ K, it
is not clear what is the source of the spatially extended cold gas. It
is still possible that the cold gas originates from within the
streams. \citet{faucher10} computed that the spectrum of
`gravitionally heated' cold streams should be double peaked and mostly
centered on the systemic velocity. This is different from what is
observed in B1. \citet{francis96} found that the \lya\ emission of B1
has an observed width of $\sim$600 km s$^{-1}$, and is redshifted by
490 km s$^{-1}$ with respect to \civ. This is similar to the typical
offset between \lya\ and interstellar absorption features (like \civ)
seen in Lyman Break Galaxies \citep{shapley03}. The redshifted \lya\
emission \footnote{We further note that \lya\ is also seen in
  absorption along the line of sight of a background QSO at an impact
  parameter of 180 kpc. The absorption is unusually broad
  ($\gtrsim1000$ km s$^{-1}$), and its line center coincides with that
  of \civ.} suggests that the photons scatter through cold gas in some
large-scale outflow \citep[as in
e.g.][]{steidel11,dijkstra12}. These outflows can explain the level of 
enrichment of the gas, which may be difficult to maintain if strong cold
flows were active. It is interesting to note that our analysis of the optical emission lines
showed that although the gas is highly turbulent (velocity dispersions
of 800-1000 km s$^{-1}$), it does not have any apparent large velocity
shear. This may hold some clues to the process of AGN feedback. The
lack of velocity shears may indicate that radiative-mode feedback from
the AGN is not capable of inducing large-scale radial flows. LABs
might thus offer good test beds for our ideas about radiative mode AGN
feedback and its impact on massive galaxy formation.

If the luminous LAB phenomenon is associated with powerful AGN, we
would naively expect that the population of LABs should extend toward
much lower redshifts than currently probed by \lya\ studies, possibly
closely following the evolution in the quasar luminosity function. In
contrast, \citet{keel09} have suggested that LABs may be specific to
the high redshift universe, based on their absence in galaxy clusters
at $z\sim0.8$. Their study was motivated by the empirical relation
between LABs and overdense environments at $z>2$. However, the highly
active and gas-rich environments typical of overdense environments at
$z>2$ (``protoclusters'') are far from equivalent to the cores of
massive, virialized clusters at $z<1$ that are dominated by large
amounts of hot gas. Instead, we propose that luminous LABs at low
redshift may still be found in the more typical, lower density, and
gas-rich environments of actively accreting galaxies and AGN. Evidence
for this exists in the form of extended ionized nebulae ($>$ few tens
of kpc in some cases) around $z<1$ radio galaxies and quasars (both
radio loud and radio quiet). Although we do not have observations of
their \lya, it is expected that they also have extended \lya\ nebulae,
possibly more extended than \oiii\ or \ha\ given that it is an
intrinsically stronger line (unless heavily absorbed).

It is also possible that the most luminous and extended LABs are only
found in the high redshift regime when the cosmological accretion
rates of cold gas were highest, providing the source of the gas that
is being ionized by the AGN. \citet{zirm09} showed that the total
luminosities (and sizes) of the \lya\ halos around HzRGs strongly
decline with decreasing redshift, which does not match any change in
the total energy output by the AGN as inferred from their radio or
X-ray luminosities. The evolution of their \lya\ halos thus appears to
be probing a real evolutionary effect perhaps related to the reservoir
of gas from which the massive galaxies are forming. This indicates
that LABs could still play an important role in the study of
cosmological gas accretion modes, although perhaps in a more indirect
manner than typically proposed.

\section{Summary}

We have used the SINFONI integral field spectrograph to resolve the
dominant rest-frame optical emission lines of the luminous
($L_{Ly\alpha}=8\times10^{43}$ erg s$^{-1}$) \lya\ blob `B1' at
$z=2.38$ discovered by \citet{francis96}. Our main findings and
conclusions are as follows.

\smallskip $\bullet$ We detect luminous
\oiii$\lambda\lambda4959,5007$\AA\ and \ha\ emission with a spatial
extent of at least $32\times40$ kpc ($4\arcsec\times5\arcsec$). The
dominant optical emission line component (B1 North) shows relatively
broad lines (600--800 km s$^{-1}$, FWHM) and AGN-photoionized
line-ratios in an optical line diagnostic diagram of \oiii/\hb\ versus
\nii/\ha.  A secondary component (B1 South) of much more modest
velocities of $\sim200$ km s$^{-1}$ (FWHM) roughly coincides with a
filament of faint \lya\ and UV continuum detected previously
\citep{francis96}.

\smallskip $\bullet$ The extinction-corrected \oiii$\lambda$5007
luminosity in B1 is about 1$\times10^{44}$ erg s$^{-1}$. This is high
compared to that of typical LAEs, but consistent with that of HzRGs
\citep{humphrey08} and quasars \citep[e.g.][]{kim13} at that
redshift. The high luminosity, combined with the evidence for AGN
photo-ionization, suggests that B1 North is the site of a hidden
quasar. This is confirmed by the fact that \oii\ is relatively weak
compared to \oiii\ (extinction-corrected \oiii/\oii\ of about 3.8),
which is indicative of a high, Seyfert-like ionization parameter. This
is consistent with the previous detection of a narrow
\civ$\lambda$1549\AA\ line, as well as with the high mid-IR luminosity
found by \citet{colbert06,colbert11}. Although the \civ\ emission was
originally interpreted as evidence of an AGN \citep{francis96}, more
recently it was hypothesized that the \civ\ and \lya\ emission could
arise from a great number of merging halos and shocked gas clouds
\citep{francis12}. In light of the new evidence, we adopt the former
interpretation, and conclude that the B1 emission line nebula is the
narrow line region of an obscured quasar.

\smallskip $\bullet$ Based on established relations between
$L_{[OIII]}$\ and $L_{bol,agn}$ for Type II AGN
\citep{heckman04,lamastra09} we conclude that B1 North hosts an
extremely luminous quasar with a bolometric luminosity of
$\sim3\times10^{46}$ erg s$^{-1}$. The obscured AGN may be
Compton-thick given existing X-ray limits.

\smallskip $\bullet$ We have performed a census of the most luminous
LABs selected from the literature, and find that virtually all 
luminous LABs ($L_{Ly\alpha}\gtrsim5\times10^{43}$ erg s$^{-1}$)
harbor obscured quasars. The properties of LABs in general are
furthermore remarkably similar to those of HzRGs and quasars.

\smallskip $\bullet$ We find that the AGN scenario is easily capable
of producing sufficient ionizing photons to power the \lya\
luminosities observed.

\smallskip $\bullet$ The fact that the duty cycle of AGN is relatively
short compared to that of cosmological gas accretion as predicted by
models, implies that AGN are the main driver of the \lya\ luminosity
in the most luminous LABs, even if the large-scale cosmological gas
flows are providing the material.

\smallskip $\bullet$ Our findings also imply that the empirical
relation between LABs and overdense environments at high redshift
suggested by the literature, is primarily due to a more fundamental
correlation between AGN (or massive galaxies) and environment.

\section*{Acknowledgments}

We thank Lee Armus, Yi-Kuan Chiang, Carlos De Breuck, Paul Francis, Tim Heckman,
Yuichi Matsuda, Emily McLinden, Masami Ouchi, Tomoki Saito, Jingwen
Wu, and the anonymous referee for comments, suggestions, and answering our questions.


\begin{thebibliography}{}
\bibitem[Alexander et al.(2005)]{alexander05} Alexander, D.~M., Bauer, F.~E., Chapman, S.~C., et al.\ 2005, \apj, 632, 736 
\bibitem[Allen et al.(2008)]{allen08} Allen, M.~G., Groves, B.~A., Dopita, M.~A., Sutherland, R.~S., 
\& Kewley, L.~J.\ 2008, \apjs, 178, 20 
\bibitem[Baldwin et al.(1981)]{baldwin81} Baldwin, J.~A., Phillips, M.~M., \& Terlevich, R.\ 1981, \pasp, 93, 5 
\bibitem[Barger et al.(2002)]{barger02} Barger, A.~J., Cowie, 
L.~L., Brandt, W.~N., et al.\ 2002, \aj, 124, 1839 
\bibitem[Basu-Zych \& Scharf(2004)]{basu-zych04} Basu-Zych, A., \& Scharf, C.\ 2004, \apjl, 615, L85 
\bibitem[Baum \& Heckman(1989)]{baum89} Baum, S.~A., \& Heckman, T.\ 1989, \apj, 336, 681 
\bibitem[Beelen et al.(2008)]{beelen08} Beelen, A., et al.\ 2008, \aap, 485, 645 
\bibitem[Best et al.(2000)]{best00} Best, P.~N., R{\"o}ttgering, H.~J.~A., \& Longair, M.~S.\ 2000, \mnras, 311, 23 
\bibitem[Bower et al.(2004)]{bower04} Bower, R.~G., Morris, S.~L., Bacon, R., et al.\ 2004, \mnras, 351, 63 
\bibitem[Bridge et al.(2012)]{bridge12} Bridge, C.~R., Blain, A., Borys, C.~J.~K., et al.\ 2012, arXiv:1205.4030 
\bibitem[Brinchmann et al.(2008)]{brinchmann08} Brinchmann, J., Pettini, M., \& Charlot, S.\ 2008, \mnras, 385, 769 
\bibitem[Chapman et al.(2003)]{chapman03} Chapman, S.~C., Blain, A.~W., Ivison, R.~J., \& Smail, I.~R.\ 2003, \nat, 422, 695 
\bibitem[Chapman et al.(2004)]{chapman04} Chapman, S.~C., Scott, D., Windhorst, R.~A., et al.\ 2004, \apj, 606, 85 
\bibitem[Chary \& Elbaz(2001)]{chary01} Chary, R., \& Elbaz, D.\ 2001, \apj, 556, 562 
\bibitem[Christensen et al.(2006)]{christensen06} Christensen, L., Jahnke, K., Wisotzki, L., \& S{\'a}nchez, S.~F.\ 2006, \aap, 459, 717 
\bibitem[Colbert et al.(2006)]{colbert06} Colbert, J.~W., Teplitz, H., Francis, P., Palunas, P., Williger, G.~M., \& Woodgate, B.\ 2006, \apjl, 637, L89 
\bibitem[Colbert et al.(2008)]{colbert08} Colbert, J.~W., Teplitz, H., Francis, P., Palunas, P., Williger, G.~M., \& Woodgate, B.\ 2008, Infrared Diagnostics of Galaxy Evolution, 381, 468 
\bibitem[Colbert et al.(2011)]{colbert11} Colbert, J.~W., Scarlata, C., Teplitz, H., et al.\ 2011, \apj, 728, 59 
\bibitem[Dey et al.(2005)]{dey05} Dey, A., Bian, C., Soifer, B.~T., et al.\ 2005, \apj, 629, 654 
\bibitem[Dijkstra \& Loeb(2009)]{dijkstra09} Dijkstra, M., \& Loeb, A.\ 2009, \mnras, 400, 1109 
\bibitem[Dijkstra \& Kramer(2012)]{dijkstra12} Dijkstra, M., \& Kramer, R.\ 2012, \mnras, 424, 1672 
\bibitem[Dopita \& Sutherland(1995)]{dopita95} Dopita, M.~A., \& Sutherland, R.~S.\ 1995, \apj, 455, 468 
\bibitem[Erb et al.(2006)]{erb06} Erb, D.~K., Shapley, A.~E., Pettini, M., et al.\ 2006, \apj, 644, 813 
\bibitem[Erb et al.(2011)]{erb11} Erb, D.~K., Bogosavljevi{\'c}, M., \& Steidel, C.~C.\ 2011, \apjl, 740, L31 
\bibitem[Fardal et al.(2001)]{fardal01} Fardal, M.~A., Katz, N., Gardner, J.~P., et al.\ 2001, \apj, 562, 605 
\bibitem[Faucher-Gigu{\`e}re et al.(2010)]{faucher10} Faucher-Gigu{\`e}re, C.-A., Kere{\v s}, D., Dijkstra, M., Hernquist, L., 
\& Zaldarriaga, M.\ 2010, \apj, 725, 633 
\bibitem[F{\"o}rster Schreiber et al.(2006)]{forster06} F{\"o}rster Schreiber, N.~M., Genzel, R., Lehnert, M.~D., et al.\ 2006, \apj, 645, 1062 
\bibitem[Francis et al.(1996)]{francis96} Francis, P.~J., et al.\ 1996, \apj, 457, 490 
\bibitem[Francis et al.(1997)]{francis97} Francis, P.~J., Woodgate, B.~E., \& Danks, A.~C.\ 1997, \apjl, 482, L25 
\bibitem[Francis et al.(2001)]{francis01} Francis, P.~J., et al.\ 2001, \apj, 554, 1001 
\bibitem[Francis et al.(2004a)]{francis04a} Francis, P.~J., Palunas, P., Teplitz, H.~I., Williger, G.~M., \& Woodgate, B.~E.\ 2004a, \apj, 614, 75 
\bibitem[Francis \& Williger(2004b)]{francis04b} Francis, P.~J., \& Williger, G.~M.\ 2004b, \apjl, 602, L77 
\bibitem[Francis et al.(2012)]{francis12} Francis, P.~J., Dopita, M.~A., Colbert, J.~W., et al.\ 2012, \mnras, 6 
\bibitem[Fu \& Stockton(2009)]{fu09} Fu, H., \& Stockton, A.\ 2009, \apj, 690, 953 
\bibitem[Geach et al.(2005)]{geach05} Geach, J.~E., Matsuda, Y., Smail, I., et al.\ 2005, \mnras, 363, 1398 
\bibitem[Geach et al.(2009)]{geach09} Geach, J.~E., Alexander, D.~M., Lehmer, B.~D., et al.\ 2009, \apj, 700, 1 
\bibitem[Gould \& Weinberg(1996)]{gould96} Gould, A., \& Weinberg, D.~H.\ 1996, \apj, 468, 462 
\bibitem[Harrison et al.(2012)]{harrison12} Harrison, C.~M., Alexander, D.~M., Swinbank, A.~M., et al.\ 2012, \mnras, 426, 1073 
\bibitem[Hatch et al.(2008)]{hatch08} Hatch, N.~A., Overzier, R.~A., R{\"o}ttgering, H.~J.~A., Kurk, J.~D., \& Miley, G.~K.\ 2008, \mnras, 383, 931 
\bibitem[Hatch et al.(2009)]{hatch09} Hatch, N.~A., Overzier, R.~A., Kurk, J.~D., et al.\ 2009, \mnras, 395, 114 
\bibitem[Hayashi et al.(2012)]{hayashi12} Hayashi, M., Kodama, T., Tadaki, K.-i., Koyama, Y., \& Tanaka, I.\ 2012, \apj, 757, 15 
\bibitem[Heckman et al.(1982)]{heckman82} Heckman, T.~M., Miley, G.~K., Balick, B., van Breugel, W.~J.~M., \& Butcher, H.~R.\ 1982, \apj, 262, 529 
\bibitem[Heckman et al.(1987)]{1987AJ.....93..276H} Heckman, T.~M., Armus, L., \& Miley, G.~K.\ 1987, \aj, 93, 276 
\bibitem[Heckman et al.(1991)]{heckman91} Heckman, T.~M., Miley, G.~K., Lehnert, M.~D., \& van Breugel, W.\ 1991, \apj, 370, 78 
\bibitem[Heckman et al.(2004)]{heckman04} Heckman, T.~M., 
Kauffmann, G., Brinchmann, J., et al.\ 2004, \apj, 613, 109 
\bibitem[Humphrey et al.(2008)]{humphrey08} Humphrey, A., Villar-Mart{\'{\i}}n, M., Vernet, J., et al.\ 2008, \mnras, 383, 11 
\bibitem[Kauffmann et al.(2003)]{kauffmann03} Kauffmann, G., Heckman, T.~M., Tremonti, C., et al.\ 2003, \mnras, 346, 1055 
\bibitem[Keel et al.(1999)]{keel99} Keel, W.~C., Cohen, S.~H., Windhorst, R.~A., \& Waddington, I.\ 1999, \aj, 118, 2547 
\bibitem[Keel et al.(2002)]{keel02} Keel, W.~C., Wu, W., Waddington, I., Windhorst, R.~A., \& Pascarelle, S.~M.\ 2002, \aj, 123, 3041 
\bibitem[Keel et al.(2009)]{keel09} Keel, W.~C., White, R.~E., III, Chapman, S., \& Windhorst, R.~A.\ 2009, \aj, 138, 986 
\bibitem[Keel et al.(2012)]{keel12} Keel, W.~C., Chojnowski, S.~D., Bennert, V.~N., et al.\ 2012, \mnras, 420, 878 
\bibitem[Kennicutt(1998)]{kennicutt98} Kennicutt, R.~C., Jr.\ 1998, \araa, 36, 189 
\bibitem[Kewley et al.(2001)]{kewley01} Kewley, L.~J., Dopita, M.~A., Sutherland, R.~S., Heisler, C.~A., \& Trevena, J.\ 2001, \apj, 556, 121 
\bibitem[Kewley et al.(2006)]{kewley06} Kewley, L.~J., Groves, B., Kauffmann, G., \& Heckman, T.\ 2006, \mnras, 372, 961 
\bibitem[Kim et al.(2013)]{kim13} Kim, M., Ho, L.~C., Lonsdale, C.~J., et al.\ 2013, arXiv:1303.7194 
\bibitem[Kriek et al.(2007)]{kriek07} Kriek, M., van Dokkum, P.~G., Franx, M., et al.\ 2007, \apj, 669, 776 
\bibitem[Kuiper et al.(2011)]{kuiper11} Kuiper, E., Hatch, 
N.~A., Miley, G.~K., et al.\ 2011, \mnras, 415, 2245 
\bibitem[Lamastra et al.(2009)]{lamastra09} Lamastra, A., Bianchi, S., Matt, G., et al.\ 2009, \aap, 504, 73 
\bibitem[Lehnert \& Heckman(1994)]{lehnert94} Lehnert, M.~D., \& Heckman, T.~M.\ 1994, \apjl, 426, L27 
\bibitem[Lehnert et al.(2009)]{lehnert09} Lehnert, M.~D., 
Nesvadba, N.~P.~H., Le Tiran, L., et al.\ 2009, \apj, 699, 1660 
\bibitem[Leitherer et al.(1999)]{leitherer99} Leitherer, C., Schaerer, D., Goldader, J.~D., et al.\ 1999, \apjs, 123, 3 
\bibitem[Le Tiran et al.(2011)]{letiran11} Le Tiran, L., Lehnert, M.~D., Di Matteo, P., Nesvadba, N.~P.~H., \& van Driel, W.\ 2011, \aap, 530, L6 
\bibitem[Maschietto et al.(2008)]{maschietto08} Maschietto, F., 
Hatch, N.~A., Venemans, B.~P., et al.\ 2008, \mnras, 389, 1223 
\bibitem[Matsuda et al.(2004)]{matsuda04} Matsuda, Y., Yamada, T., Hayashino, T., et al.\ 2004, \aj, 128, 569 
\bibitem[Matsuda et al.(2006)]{matsuda06} Matsuda, Y., Yamada, T., Hayashino, T., Yamauchi, R., \& Nakamura, Y.\ 2006, \apjl, 640, L123 
\bibitem[Matsuda et al.(2009)]{matsuda09} Matsuda, Y., Nakamura, Y., Morimoto, N., et al.\ 2009, \mnras, 400, L66 
\bibitem[Matsuda et al.(2011)]{matsuda11} Matsuda, Y., Yamada, T., Hayashino, T., et al.\ 2011, \mnras, 410, L13 
\bibitem[McCarthy et al.(1987)]{mccarthy89} McCarthy, P.~J., Spinrad, H., Djorgovski, S., et al.\ 1987, \apjl, 319, L39 
\bibitem[McCarthy et al.(1990)]{mccarthy90} McCarthy, P.~J., 
Spinrad, H., Dickinson, M., et al.\ 1990, \apj, 365, 487 
\bibitem[McLinden et al.(2011)]{mclinden11} McLinden, E.~M., 
Finkelstein, S.~L., Rhoads, J.~E., et al.\ 2011, \apj, 730, 136 
\bibitem[Men{\'e}ndez-Delmestre et al.(2009)]{menendez-delmestre09} Men{\'e}ndez-Delmestre, K., Blain, A.~W., Smail, I., et al.\ 2009, \apj, 
699, 667 
\bibitem[Miley et al.(2004)]{miley04} Miley, G.~K., Overzier, R.~A., Tsvetanov, Z.~I., et al.\ 2004, Nature, 427, 47 
\bibitem[Miley et al.(2006)]{miley06} Miley, G.~K., Overzier, R.~A., Zirm, A.~W., et al.\ 2006, \apjl, 650, L29 
\bibitem[Nakajima et al.(2012)]{nakajima12} Nakajima, K., Ouchi, M., Shimasaku, K., et al.\ 2012, arXiv:1208.3260 
\bibitem[Nesvadba et al.(2006)]{nesvadba06} Nesvadba, N.~P.~H., Lehnert, M.~D., Eisenhauer, F., et al.\ 2006, \apj, 650, 693 
\bibitem[Nesvadba et al.(2007a)]{nesvadba07a} Nesvadba, N.~P.~H., Lehnert, M.~D., De Breuck, C., Gilbert, A., \& van Breugel, W.\ 2007a, \aap, 475, 145 
\bibitem[Nesvadba et al.(2008)]{nesvadba08} Nesvadba, N.~P.~H., Lehnert, M.~D., Davies, R.~I., Verma, A., \& Eisenhauer, F.\ 2008, \aap, 479, 67 
\bibitem[Nesvadba et al.(2011)]{nesvadba11} Nesvadba, N.~P.~H., Polletta, M., Lehnert, M.~D., et al.\ 2011, \mnras, 415, 2359 
\bibitem[Netzer et al.(2004)]{netzer04} Netzer, H., Shemmer, O., Maiolino, R., et al.\ 2004, \apj, 614, 558 
\bibitem[Nilsson et al.(2006)]{nilsson06} Nilsson, K.~K., Fynbo, J.~P.~U., M{\o}ller, P., Sommer-Larsen, J., \& Ledoux, C.\ 2006, \aap, 452, L23 
\bibitem[North et al.(2012)]{north12} North, P.~L., Courbin, F., Eigenbrod, A., \& Chelouche, D.\ 2012, \aap, 542, A91 
\bibitem[Overzier et al.(2001)]{overzier01} Overzier, R.~A., R{\"o}ttgering, H.~J.~A., Kurk, J.~D., \& De Breuck, C.\ 2001, \aap, 370, L39
\bibitem[Overzier et al.(2006)]{overzier06} Overzier, R.~A., Miley, G.~K., Bouwens, R.~J., et al.\ 2006, \apj, 637, 58 
\bibitem[Overzier et al.(2008)]{overzier08} Overzier, R.~A., Bouwens, R.~J., Cross, N.~J.~G., et al.\ 2008, \apj, 673, 143 
\bibitem[Overzier et al.(2009)]{overzier09} Overzier, R.~A., 
Heckman, T.~M., Tremonti, C., et al.\ 2009, \apj, 706, 203 
\bibitem[Palunas et al.(2004)]{palunas04} Palunas, P., Teplitz, H.~I., Francis, P.~J., Williger, G.~M., 
\& Woodgate, B.~E.\ 2004, \apj, 602, 545 
\bibitem[Pascarelle et al.(1996)]{pascarelle96} Pascarelle, S.~M., 
Windhorst, R.~A., Keel, W.~C., \& Odewahn, S.~C.\ 1996, \nat, 383, 45 
\bibitem[Pentericci et al.(1998)]{pentericci98} Pentericci, L., Roettgering, H.~J.~A., Miley, G.~K., et al.\ 1998, \apj, 504, 139 
\bibitem[Pentericci et al.(1999)]{pentericci99} Pentericci, L., R{\"o}ttgering, H.~J.~A., Miley, G.~K., et al.\ 1999, \aap, 341, 329 
\bibitem[Pentericci et al.(2000)]{pentericci00} Pentericci, L., Kurk, J.~D., R{\"o}ttgering, H.~J.~A., et al.\ 2000, \aap, 361, L25 
\bibitem[Pettini et al.(2001)]{pettini01} Pettini, M., Shapley, 
A.~E., Steidel, C.~C., et al.\ 2001, \apj, 554, 981 
\bibitem[Polletta et al.(2008)]{polletta08} Polletta, M., Omont, A., Berta, S., et al.\ 2008, \aap, 492, 81 
\bibitem[Polletta et al.(2011)]{polletta11} Polletta, M., Nesvadba, N.~P.~H., Neri, R., et al.\ 2011, \aap, 533, A20 
\bibitem[Prescott et al.(2008)]{prescott08} Prescott, M.~K.~M., Kashikawa, N., Dey, A., \& Matsuda, Y.\ 2008, \apjl, 678, L77 
\bibitem[Prescott et al.(2009)]{prescott09} Prescott, M.~K.~M., Dey, A., \& Jannuzi, B.~T.\ 2009, \apj, 702, 554 
\bibitem[Prescott et al.(2012)]{prescott12} Prescott, M.~K.~M., Dey, A., Brodwin, M., et al.\ 2012, \apj, 752, 86 
\bibitem[Reuland et al.(2003)]{reuland03} Reuland, M., van 
Breugel, W., R{\"o}ttgering, H., et al.\ 2003, \apj, 592, 755 
\bibitem[Rich et al.(2011)]{rich11} Rich, J.~A., Kewley, L.~J., \& Dopita, M.~A.\ 2011, \apj, 734, 87 
\bibitem[Richards et al.(2006)]{richards06} Richards, G.~T., Lacy, M., Storrie-Lombardi, L.~J., et al.\ 2006, \apjs, 166, 470 
\bibitem[Saito et al.(2006)]{saito06} Saito, T., Shimasaku, K., Okamura, S., et al.\ 2006, \apj, 648, 54 
\bibitem[Sanders \& Mirabel(1996)]{sanders96} Sanders, D.~B., \& Mirabel, I.~F.\ 1996, \araa, 34, 749 
\bibitem[Scarlata et al.(2009)]{scarlata09} Scarlata, C., Colbert, J., Teplitz, H.~I., et al.\ 2009, \apj, 706, 1241 
\bibitem[Shapley et al.(2003)]{shapley03} Shapley, A.~E., Steidel, C.~C., Pettini, M., \& Adelberger, K.~L.\ 2003, \apj, 588, 65 
\bibitem[Shen et al.(2007)]{shen07} Shen, Y., Strauss, M.~A., 
Oguri, M., et al.\ 2007, \aj, 133, 2222 
\bibitem[Smith \& Jarvis(2007)]{smith07} Smith, D.~J.~B., \& Jarvis, M.~J.\ 2007, \mnras, 378, L49 
\bibitem[Smith et al.(2008)]{smith08} Smith, D.~J.~B., Jarvis, M.~J., Lacy, M., \& Mart{\'{\i}}nez-Sansigre, A.\ 2008, \mnras, 389, 799 
\bibitem[Smith et al.(2009)]{smith09} Smith, D.~J.~B., Jarvis, M.~J., Simpson, C., \& Mart{\'{\i}}nez-Sansigre, A.\ 2009, \mnras, 393, 309 
\bibitem[Sol{\'o}rzano-I{\~n}arrea et al.(2001)]{solorzano01} Sol{\'o}rzano-I{\~n}arrea, C., Tadhunter, C.~N., 
\& Axon, D.~J.\ 2001, \mnras, 323, 965 
\bibitem[Steidel et al.(2000)]{steidel00} Steidel, C.~C., Adelberger, K.~L., Shapley, A.~E., et al.\ 2000, \apj, 532, 170 
\bibitem[Steidel et al.(2011)]{steidel11} Steidel, C.~C., Bogosavljevi{\'c}, M., Shapley, A.~E., et al.\ 2011, \apj, 736, 160 
\bibitem[Tadhunter et al.(1986)]{tadhunter86} Tadhunter, C.~N., Perez, E., \& Fosbury, R.~A.~E.\ 1986, \mnras, 219, 555 
\bibitem[Teplitz et al.(2000)]{teplitz00} Teplitz, H.~I., McLean, I.~S., Becklin, E.~E., et al.\ 2000, \apjl, 533, L65 
\bibitem[van Dokkum et al.(2005)]{vandokkum05} van Dokkum, P.~G., Kriek, M., Rodgers, B., Franx, M., \& Puxley, P.\ 2005, \apjl, 622, L13 
\bibitem[Veilleux \& Osterbrock(1987)]{veilleux87} Veilleux, S., \& Osterbrock, D.~E.\ 1987, \apjs, 63, 295 
\bibitem[Venemans et al.(2002)]{venemans02} Venemans, B.~P., Kurk, J.~D., Miley, G.~K., et al.\ 2002, \apjl, 569, L11 
\bibitem[Venemans et al.(2005)]{venemans05} Venemans, B.~P., R{\"o}ttgering, H.~J.~A., Miley, G.~K., et al.\ 2005, \aap, 431, 793 
\bibitem[Venemans et al.(2007)]{venemans07} Venemans, B.~P., R{\"o}ttgering, H.~J.~A., Miley, G.~K., et al.\ 2007, \aap, 461, 823 
\bibitem[Villar-Martin et al.(1997)]{villar-martin97} Villar-Martin, M., Tadhunter, C., \& Clark, N.\ 1997, \aap, 323, 21 
\bibitem[Villar-Mart{\'{\i}}n et al.(1999)]{villar-martin99} Villar-Mart{\'{\i}}n, M., Binette, L., \& Fosbury, R.~A.~E.\ 1999, \aap, 346, 7 
\bibitem[Villar-Mart{\'{\i}}n et al.(2000)]{villar-martin00} Villar-Mart{\'{\i}}n, M., Alonso-Herrero, A., di Serego Alighieri, S., \& Vernet, J.\ 2000, \aaps, 147, 291 
\bibitem[Villar-Mart{\'{\i}}n et al.(2002)]{villar-martin02} Villar-Mart{\'{\i}}n, M., Vernet, J., di Serego Alighieri, S., et al.\ 
2002, \mnras, 336, 436 
\bibitem[Villar-Mart{\'{\i}}n et al.(2003)]{villar-martin03} Villar-Mart{\'{\i}}n, M., Vernet, J., di Serego Alighieri, S., et al.\ 
2003, \mnras, 346, 273 
\bibitem[Villar-Mart{\'{\i}}n et al.(2006)]{villar-martin06} Villar-Mart{\'{\i}}n, M., S{\'a}nchez, S.~F., De Breuck, C., et al.\ 2006, \mnras, 366, L1 
\bibitem[Villar-Mart{\'{\i}}n et al.(2007a)]{villar-martin07a}Villar-Mart{\'{\i}}n, M., Humphrey, A., De Breuck, C., et al.\ 2007a, 
\mnras, 375, 1299 
\bibitem[Villar-Mart{\'{\i}}n et al.(2007b)]{villar-martin07b}Villar-Mart{\'{\i}}n, M., S{\'a}nchez, S.~F., Humphrey, A., et al.\ 2007b, 
\mnras, 378, 416 
\bibitem[Villar-Mart{\'{\i}}n et al.(2011)]{villar-martin11} Villar-Mart{\'{\i}}n, M., Humphrey, A., Delgado, R.~G., Colina, L., \& Arribas, S.\ 2011, \mnras, 418, 2032 
\bibitem[Webb et al.(2009)]{webb09} Webb, T.~M.~A., Yamada, T., Huang, J.-S., et al.\ 2009, \apj, 692, 1561 
\bibitem[Weijmans et al.(2010)]{weijmans10} Weijmans, A.-M., Bower, R.~G., Geach, J.~E., et al.\ 2010, \mnras, 402, 2245 
\bibitem[Wilman et al.(2005)]{wilman05} Wilman, R.~J., Gerssen, J., Bower, R.~G., Morris, S.~L., Bacon, R., de Zeeuw, P.~T., \& Davies, R.~L.\ 2005, \nat, 436, 227 
\bibitem[Wu et al.(2010)]{wu10} Wu, Y., Helou, G., Armus, L., et al.\ 2010, \apj, 723, 895 
\bibitem[Wu et al.(2012)]{wu12} Wu, J., Tsai, C.-W., Sayers, J., et al.\ 2012, \apj, 756, 96 
\bibitem[Yang et al.(2009)]{yang09} Yang, Y., Zabludoff, A., Tremonti, C., Eisenstein, D., \& Dav{\'e}, R.\ 2009, \apj, 693, 1579 
\bibitem[Yang et al.(2010)]{yang10} Yang, Y., Zabludoff, A., Eisenstein, D., \& Dav{\'e}, R.\ 2010, \apj, 719, 1654 
\bibitem[Yang et al.(2011)]{yang11} Yang, Y., Zabludoff, A., Jahnke, K., et al.\ 2011, \apj, 735, 87 
\bibitem[Zirm et al.(2005)]{zirm05} Zirm, A.~W., Overzier, R.~A., Miley, G.~K., et al.\ 2005, \apj, 630, 68 
\bibitem[Zirm et al.(2009)]{zirm09} Zirm, A.~W., Dey, A., Dickinson, M., \& Norman, C.~J.\ 2009, \apjl, 694, L31 
\end{thebibliography}
\end{document}